\documentclass{article}

\usepackage{microtype}
\usepackage{graphicx}
\usepackage{subfigure}
\usepackage{booktabs} 

\usepackage{hyperref}

\usepackage[accepted]{icml2024}

\usepackage{amsmath}
\usepackage{amssymb}
\usepackage{mathtools}
\usepackage{amsthm}

\usepackage[capitalize,noabbrev]{cleveref}

\usepackage{bm}
\usepackage{multirow}
\usepackage{multicol}
\usepackage{makecell}
\usepackage{balance}
\theoremstyle{plain}

\theoremstyle{definition}

\theoremstyle{remark}

\usepackage[textsize=tiny]{todonotes}
\usepackage{pifont}
\usepackage{xcolor}
\usepackage{graphicx}
\usepackage{subcaption}
\usepackage{caption}
\usepackage{mdframed}
\begin{document}

\setlength{\abovedisplayskip}{3pt}
\setlength{\belowdisplayskip}{3pt}
\setlength{\abovedisplayshortskip}{3pt}
\setlength{\belowdisplayshortskip}{3pt}

\twocolumn[
\icmltitle{LangCell: Language-Cell Pre-training for Cell Identity Understanding}



\icmlsetsymbol{equal}{*}
\icmlsetsymbol{corr}{\dag}

\begin{icmlauthorlist}
\icmlauthor{Suyuan Zhao}{equal,air,thu}
\icmlauthor{Jiahuan Zhang}{equal,air}
\icmlauthor{Yushuai Wu}{air}
\icmlauthor{Yizhen Luo}{air,thu}
\icmlauthor{Zaiqing Nie}{corr,air,phar}
\end{icmlauthorlist}

\icmlaffiliation{thu}{Department of Computer Science and Tecnology, Tsinghua University}
\icmlaffiliation{air}{Institute for AI Industry Research (AIR), Tsinghua University}
\icmlaffiliation{phar}{PharMolix Inc.}

\icmlcorrespondingauthor{Suyuan Zhao}{sxdtzsy@gmail.com}
\icmlcorrespondingauthor{Jiahuan Zhang}{zhangjiahuan@air.tsinghua.edu.cn}
\icmlcorrespondingauthor{Zaiqing Nie}{zaiqing@air.tsinghua.edu.cn}

\icmlkeywords{Machine Learning, ICML}

\vskip 0.3in
]
\newcommand{\jh}[1]{\textcolor{blue}{\small{\bf [jh: #1]}}}

\hypersetup{
colorlinks=true,
linkcolor=blue
}


\printAffiliationsAndNotice{\icmlEqualContribution \icmlCorr} 

\begin{abstract}
Cell identity encompasses various semantic aspects of a cell, including cell type, pathway information, disease information, and more, which are essential for biologists to gain insights into its biological characteristics. Understanding cell identity from the transcriptomic data, such as annotating cell types, has become an important task in bioinformatics. 
As these semantic aspects are determined by human experts, it is impossible for AI models to effectively carry out cell identity understanding tasks without the supervision signals provided by single-cell and label pairs. 
The single-cell pre-trained language models (PLMs) currently used for this task are trained only on a single modality, transcriptomics data, lack an understanding of cell identity knowledge. As a result, they have to be fine-tuned for downstream tasks and struggle when lacking labeled data with the desired semantic labels.
To address this issue, we propose an innovative solution by constructing a unified representation of single-cell data and natural language during the pre-training phase, allowing the model to directly incorporate insights related to cell identity.
More specifically, we introduce \textbf{LangCell}, the first \textbf{Lang}uage-\textbf{Cell} pre-training framework. 
LangCell utilizes texts enriched with cell identity information to gain a profound comprehension of cross-modal knowledge.
Results from experiments conducted on different benchmarks show that LangCell is the only single-cell PLM that can work effectively in zero-shot cell identity understanding scenarios, and also significantly outperforms existing models in few-shot and fine-tuning cell identity understanding scenarios.

\end{abstract}

\section{Introduction}

Single-cell RNA sequencing (scRNA-seq) data represents a powerful tool for deciphering the ``language of life'', offering profound insights into downstream biomedical applications \cite{ziegenhain2017comparative}. 
In scRNA-seq data analysis, it is crucial to understand cell identity from multiple perspectives, such as cell type, pathway information and disease information \cite{morris2019evolving,  abdolhosseini2019cell}. Tasks like cell type annotation and cell batch integration have become the cornerstone of this field \cite{luecken2019current, scib}.
 
Pre-trained language models (PLMs) have recently demonstrated success in deciphering the complex language of life \cite{mo2021multimodal, DNABERT_ji_2021}. Building on these findings, recent studies emphasize the effectiveness and feasibility of using PLMs to analyze single-cell data solely based on sequencing information \cite{ScBERT_yang_2022, geneformer, scGPT, gong2023xtrimogene}.
These models harness the transformer architecture to assimilate millions of scRNA-seq entries, refining their capabilities through fine-tuning to adeptly perform diverse downstream tasks. 
Despite these successes, current single-cell representation models face the following challenges:

(1). Current model frameworks, which rely solely on self-supervised learning methods like masked modeling, are adept at capturing gene co-expression relationships. However, due to a lack of effective utilization of human expert knowledge, they fall short in focusing on understanding cell identity when learning cell representations. This limitation restricts the model's capacity for representation learning, consequently affecting its performance in various downstream tasks.

(2). As cell identities are determined by human experts in natural language, it is impossible for existing models to effectively carry out cell identity understanding tasks without fine-tuned with single cell and text/label pairs.
Both the amount and quality of data for fine-tuning significantly impact the model's performance in specific tasks. 
However, in practical scenarios, obtaining sufficient and reliable labeled data that closely matches the downstream task is often costly. This difficulty is even more pronounced in situations such as researching new diseases or cell subtypes, where no existing data may be available \cite{zhai2023realistic}. These practical issues significantly reduce the convenience and applicability of existing models. 

BioTranslator \cite{BioTranslator} considers combining biomedical data and natural language in the current research landscape.
However, BioTranslator only used Transformer-based model and performed large-scale pre-training on natural language modality.
It did not actually pre-train on large amounts of single-cell data but instead relied on training a naive fully connected network on downstream data, which struggles to capture the richness and complexity of transcriptomic data.

We believe that encoding scRNA-seq data with high quality and aligning it with multi-perspective textual annotations can significantly enhance the comprehension between textual information and single-cell data. 
This integration equips the model with the capacity to extend its knowledge from familiar categories to novel ones, guided by semantic coherence. This approach not only enhances the model's predictive accuracy but also bolsters its applicability across diverse biomedical scenarios. We propose \textbf{LangCell}, a genuine 
\textbf{Lang}uage-\textbf{Cell} Pre-training model to seamlessly integrate the feature space of scRNA-seq data with textual information, marking significant advancements in understanding cell identity.

We constructed a cell-text dataset, \textbf{scLibrary}, containing 27.5 million scRNA-seq entries along with their descriptions. Specifically, we obtained raw scRNA-seq data and corresponding metadata from the CELLxGENE \cite{cellxgeneDiscover}. We selected eight critical aspects of cell identity that could contain essential insights, including cell type, developmental stage and disease information, to obtain as comprehensive descriptions as possible from the Open Biological and Biomedical Ontology Foundry (OBO Foundry) \cite{obo}.

Subsequently, we have transferred some key insights from the fields of NLP and CV \cite{Li2022BLIPBL, Declip, gao-simcse, what_do_sfvt}, and designed a set of multi-task cooperative pre-training methods that are effective in the cell-text domain. Specifically, we introduce four tasks during the pre-training phase. Masked Gene Modeling (MGM) and Cell-Cell Contrastive Learning (C-C), to enhance single-cell representation learning. Additionally, we employ Cell-Text Contrastive Learning (C-T) and Cell-Text Matching (CTM) to train our model to recognize the underlying links between single-cell and textual data.

\begin{figure}
    \centering
    \setlength{\abovecaptionskip}{0.2cm}
    \includegraphics[width=
    \linewidth]{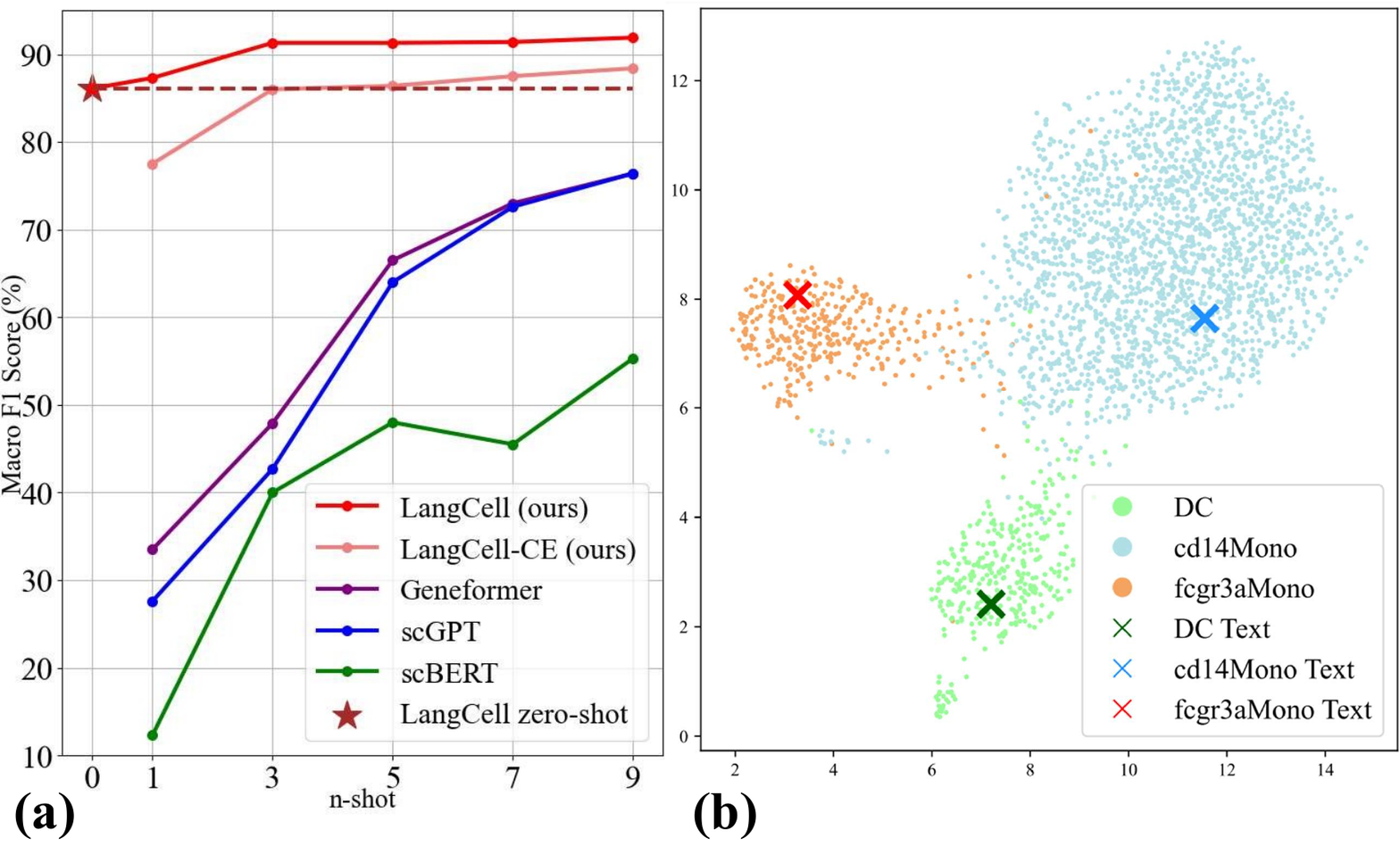}
    \caption{\textbf{(a). Plots of zero- and few-shot cell type annotation.} The curve shows the average F$_1$-scores on PBMC10K and PBMC3\&68K, for two settings of LangCell and three of the best single-cell PLMs. \quad \textbf{(b). UMAP plot of embeddings for scRNA-seq data and descriptions of three similar cell types in PBMC10K.} LangCell aligns single-cell and text embeddings. }
    \label{fig0}
    \vskip -0.2in
\end{figure}

LangCell achieves state-of-the-art (SOTA) performance on a range of cell identity understanding tasks across zero-shot, few-shot, and full dataset scenarios. It addresses classic tasks such as cell type annotation and batch integration. As the first model capable of true zero-shot cell type annotation, LangCell shows excellent performance in zero-shot scenarios, surpassing few-shot baselines in most cases (Fig. \ref{fig0}\textbf{a}).

In complex scenarios with rich cell types and low subtype distinction, we propose the cell-text retrieval task, enabling users to describe target cell types in natural language and search within databases. We also introduced two new tasks of great biological significance: non-small cell lung cancer subtype classification and pathway identification and constructed high-quality benchmarks for each. Results demonstrate that LangCell offers profound insights into cell identity from multiple perspectives, including cell type, disease subtype, and cell pathways.

Our main contributions can be summarized as follows:

(1). We introduce LangCell, the first Language-Cell pre-training framework. It unifies cellular language and natural language into a latent space. (Fig. \ref{fig0}\textbf{b}) This process infuses the model with text knowledge related to cell identity, enhancing LangCell's understanding, expression, and generalization of transcriptomic data.

(2). By harnessing the powerful link between scRNA-seq data and natural language texts, LangCell stands out as the sole PLM capable of executing zero-shot cell identity understanding tasks, surpassing the performance of existing few-shot learning models with superior experimental outcomes.

(3). LangCell's cell encoder outperforms state-of-the-art (SOTA) models in all few-shot and fine-tuning tasks related to cell identity understanding. These advancements are attributed to the synergistic impact of self-supervised learning on scRNA-seq data and distant supervision based on text.

\begin{figure*}[ht]
\begin{center}
\centerline{\includegraphics[width=1\textwidth]{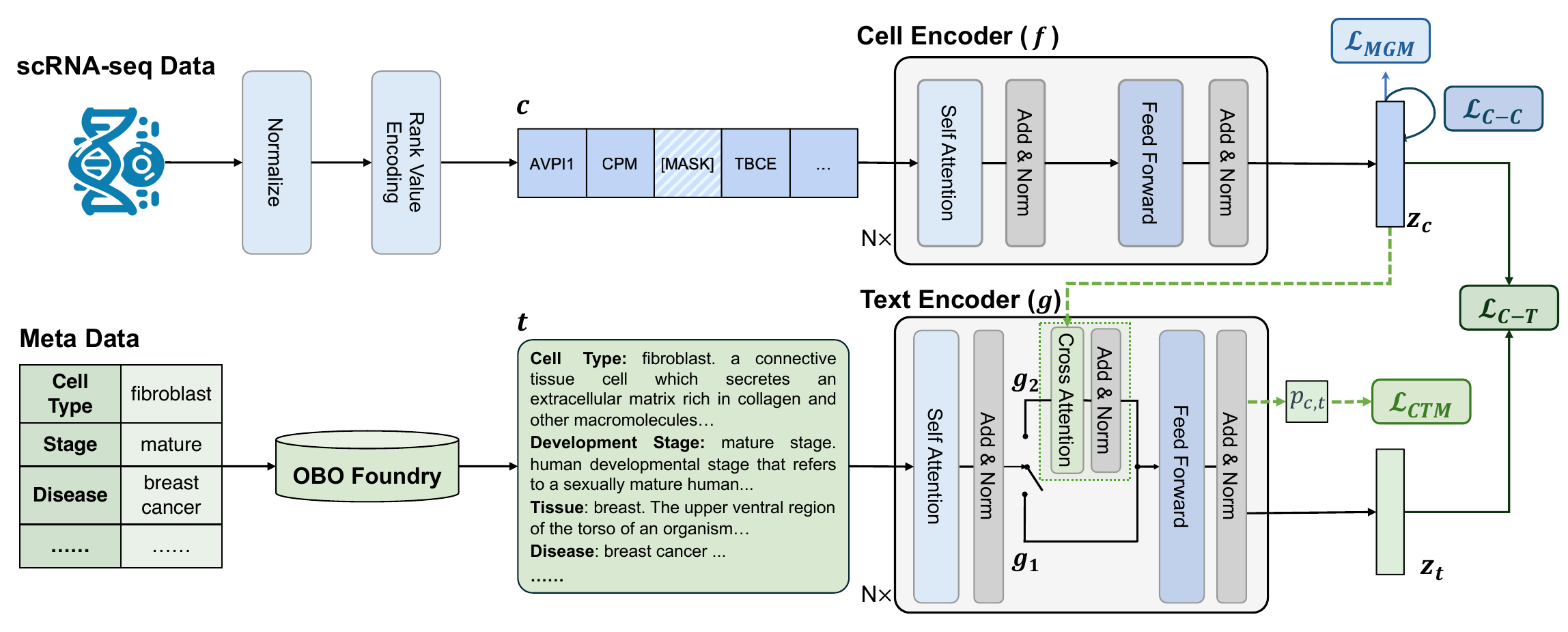}}
\caption{\textbf{The schematic overview of LangCell.} For each single-cell data containing a pair of scRNA-seq data and metadata: \textbf{(1)} During preprocessing, the scRNA-seq data is converted into a gene sequence arranged in descending order of relative expression levels, and a multi-perspective textual description of the cell is obtained from the metadata using OBO Foundry. \textbf{(2)} The embeddings of the cell and text are obtained using the cell encoder ($\bm{f}$) and the unimodal mode ($\bm{g_1}$) of the text encoder, and the matching score $p_{c,t}$ is calculated using the multimodal mode ($\bm{g_2}$) of the text encoder. \textbf{(3)} Pre-training is conducted through joint optimization of four loss functions. Among them, Masked Gene Modeling (MGM) and Cell-Cell Contrastive Learning (C-C) aim to enhance single-cell representation learning. In contrast, Cell-Text Contrastive Learning (C-T) and Cell-Text Matching (CTM) aim to train the model to understand the latent connections between single-cell and textual data.}
\label{framework}
\end{center}
\end{figure*}

\section{Related works}

\paragraph{scRNA-seq Data Representation}

PLMs offer more potential for better scRNA-seq data representation. 
scBERT \cite{ScBERT_yang_2022}, Geneformer \cite{geneformer}, scGPT \cite{scGPT}, and scFoundation \cite{scfoundation} are transformer-based models that collectively pre-train on extensive scRNA-seq datasets, ranging from over a million to 50M samples, and demonstrate advanced capabilities in tasks such as cell type annotation, transfer learning across biological tasks, drug response prediction and other tasks. BioTranslator \cite{BioTranslator} bridges the gap between natural language and scRNA-seq data. However, its reliance on MLP for encoding scRNA-seq data falls short of capturing the intricacies of transcriptomic complexity.

\paragraph{Multi-modal in Scientific Data}

Scientific data, such as molecules, proteins, and scRNA-seq data, which are not as visually intuitive as images, necessitate a more profound level of multi-modal interaction and comprehension. Works like KV-PLM \cite{KVPLM} have paved the way for a unified comprehension of molecules and textual information, while MolT5 \cite{MolT5} represents a self-supervised framework that empowers the system to handle innovative cross-modal tasks, including molecular captioning and text-based de novo molecular generation. ProtST \cite{Xu2023ProtSTML} has pioneered the field of protein multi-modal learning. However, scRNA-seq analysis has yet to witness a significant model for cross-modal representation learning, even BioTranslator has not fully realized the potential of cross-modal learning. We firmly believe that establishing connections between scRNA-seq data and text information is paramount.

\section{Methods}
In this section, we provide a comprehensive description of the LangCell workflow. The framework of the LangCell is illustrated in Fig. \ref{framework}. 
Next, we first present the processing of scRNA-seq data to fit the cell encoder in \ref{sec3.1}, then the model architecture and pre-training methods in \ref{sec3.2} and \ref{sec3.3}, respectively, and finally the downstream applications in \ref{sec3.4}.

\subsection{Data Processing}
\label{sec3.1}

The raw scRNA-seq data is provided as a count matrix. Assume there are \( m \) cells and \( n \) genes considered, then the count matrix is denoted as \( A \in \mathbb{N}^{m \times n} \). We adopted a rank value encoding method \cite{geneformer, qiu2013impact}, used for converting the count matrix into sequence data analogous to natural language. Firstly, we normalize the gene expression of each cell separately to remove the influence of sequencing depth to get \( A' \). Then, we find the non-zero median \( \beta \in \mathbb{R}^n \) for each column of \( A' \) as the median expression of each gene, and use \( \beta \) to normalize each column of \( A' \) to get \( A'' \). That is:
\[
A_{ij}' = \frac{A_{ij}}{\sum\limits_{k=1}^{n} A_{ik}}, \; 
\beta_{j} = \text{median} \{A_{kj}'|A_{kj}' \neq 0\}, \;
A_{ij}'' = \frac{A_{ij}'}{\beta_{j}}
\]
Compared to \( A' \), \( A'' \) eliminates the differences brought about by the overall expression level of the base, and its values can reflect the relative level of a gene's expression in a cell among all cells. For example, some housekeeping genes may easily have higher absolute expression, but this does not necessarily indicate a particularly noteworthy high expression of that gene in that cell. Based on \( A'' \), we sort the expressed genes in each cell by their relative expression to obtain the sequence representation of that cell. Notably, we conduct statistics on a large-scale pre-training dataset to obtain a more universal \( \beta \) and use this \( \beta \) in all subsequent model applications.

\subsection{Model Architecture}
\label{sec3.2}
Our model consists of two trainable parts: a cell encoder and a text encoder.

\textbf{Cell Encoder}: We use pre-trained Geneformer \cite{geneformer} to initialize our cell encoder, which encodes the sequential cell inputs into an embedding sequence. Notably, we add a \texttt{[CLS]} token at the beginning of the sequence, whose embedding is projected through a linear projector as a cell embedding.

\textbf{Text Encoder}: This encoder has two encoding modes: unimodal and multimodal \cite{Li2022BLIPBL}. For unimodal text encoding, it is equivalent to a BERT \cite{devlin2018bert}. For multimodal encoding, we add a pluggable cross-attention module after each self-attention module in the attention layers to compute the joint embedding and the cell-text matching probability through a linear layer. The weights are initialized using PubMedBERT \cite{pubmedbert}, which is proven to be one of the best pre-trained BERTs in the biomedical field.

Define the cell encoder as $\bm{f}$, which is utilized to derive the embedding $z_c$ from single-cell data $c$. Define the unimodal mode of the text encoder as $\bm{g_1}$, which is responsible for generating the embedding $z_t$ from textual data t. Define the multimodal mode of the text encoder as $\bm{g_2}$, which is employed to calculate the matching probability $p_{c,t}$ between single-cell and text data. These three encoding methodologies are articulated as follows:
\[
\bm{z_c} = \bm{f}(c), 
\]
\[
\; \bm{z_t} = \bm{g_1}(t), 
\]
\[
p_{c,t}=\bm{g_2}(\bm{z_c}, t)=\bm{g_2}(\bm{f}(c), t)
\]

\subsection{Pre-training Process}
\label{sec3.3}
Our model is designed to map scRNA-seq data and text to a shared latent space and utilize the unstructured knowledge contained in natural language as distant supervision to optimize cell representation learning. To this end, during the pre-training process, we jointly optimize four objective loss functions, including masked gene modeling, intra- and inter-modal contrastive learning, and cell-text matching.

\textbf{Masked Gene Modeling (MGM)}: We randomly mask some of the genes in the cell input sequence and use the output embeddings of the model at these positions to predict the reconstruction of the original input. We use the cross-entropy loss function as the loss function for this multi-class task:
\[
\mathcal{L}_{MGM}=\frac{1}{N}\sum_{i=1}^N H(v_{ij}, \hat{v}_{ij})
\]
where \(N\) is the number of masked genes,  \(v_{ij}\) and \(\hat{v}_{ij}\) respectively represent the label and predicted probability of the \(i\)-th masked position being identified as the \(j\)-th gene, $H$ is the cross entropy loss function.

\textbf{Cell-Cell Intra-Modal Contrastive Learning (C-C)}: We introduce cell-cell contrastive learning to alleviate the problem of representation degradation caused by BERT-based methods \cite{BERT_flow_li-etal-2020, Sentence-bert_reimers_gurevych_2019}. In scRNA-seq data, each gene expression level carries unique meaning, and artificial data augmentation methods such as shuffling and perturbing at the input data may disrupt the gene expression semantics \cite{Concerto_Yang2022ContrastiveLE}. We believe that perturbation at the feature level is more suitable for data augmentation in scRNA-seq data. Therefore, we use two instances of standard dropout applied to the same single-cell to construct positive samples, while other single-cells in the same batch serve as negative samples, which has been proven effective in natural language research \cite{gao-simcse}. To expand the batch size under limited video memory, we adopted a momentum encoder method similar to that of \cite{Li2022BLIPBL}. The InfoNCE \cite{MOCO_He2019} loss function is used as follows:
\[
\mathcal{L}_{C-C}=-\frac{1}{T}\sum_{i=1}^{T}\log{\frac{\mathrm{e}^{\mathrm{sim}(\bm{z_c}^{(i)}, \bm{z_c}^{(i)+})/\tau}}{\sum_{j=1}^{T}\mathrm{e}^{\mathrm{sim}(\bm{z_c}^{(i)}, \bm{z_c}^{(j)+})/\tau}}}
\]
where $T$ is the batch size, $\mathrm{sim}$ is the cosine similarity function, $\tau$ is the temperature parameter, $\bm{z_c}^{(i)}$ and $\bm{z_c}^{(i)+}$ represent the embedding of the $i$-th cell and its positive sample, respectively.

\textbf{Cell-Text Inter-Modal Contrastive Learning (C-T)}: We project cells and text into the same embedding space through cell-text contrastive learning. The text encoder employs an unimodal encoding mode. This technique has been widely used in multimodal fields such as image-text and has been proven effective \cite{clip, Li2022BLIPBL}. Similarly, a momentum encoder is used to expand the batch size. The loss function is as follows:
\begin{gather*}
    \mathcal{L}_{C-T}=-\frac{1}{2T}\sum_{i=1}^{T}(\log{\frac{\mathrm{e}^{\mathrm{sim}(\bm{z_c}^{(i)}, \bm{z_t}^{(i)})/\tau}}{\sum_{j=1}^{T}\mathrm{e}^{\mathrm{sim}(\bm{z_c}^{(i)}, \bm{z_t}^{(j)})/\tau}}} \\
    + \log{\frac{\mathrm{e}^{\mathrm{sim}(\bm{z_t}^{(i)}, \bm{z_c}^{(i)})/\tau}}{\sum_{j=1}^{T}\mathrm{e}^{\mathrm{sim}(\bm{z_t}^{(i)}, \bm{z_c}^{(j)})/\tau}}})
\end{gather*}
The symbols in the formula represent similar meanings as above. $\bm{z_t}^{(i)}$ represents the text embeddings.

\textbf{Cell-Text Matching (CTM)}: When computing this loss, the text encoder adopts a multimodal encoding mode, conducting cross-attention calculations with cell embeddings after each self-attention layer, and the final output is used for binary classification to predict whether the cell matches the text or not. This task aims to explore the matching relationship between cells and text with higher resolution, selecting cells and texts that are as similar as possible to the positive examples of cell-text pairs to form negative examples. The loss function is binary cross-entropy: 
\[ 
\mathcal{L}_{CTM}=H(y, p_{c, t})
\]
where $y$ represents the label indicating whether the cell matches the text.

\textbf{Overall Pre-training Loss}: We optimize the weighted sum of the four losses to simultaneously explore the intrinsic patterns of scRNA-seq data and its associations with text:
\[ 
\mathop{\min}\limits_{\theta} \gamma_1\mathcal{L}_{MGM}+\gamma_2\mathcal{L}_{C-C}+\gamma_3\mathcal{L}_{C-T}+\gamma_4\mathcal{L}_{CTM}
\]
where \(\theta\) represents the model parameters, and \(\gamma_i\) are the weights acting as hyperparameters.

\subsection{Downstream Applications}
\label{sec3.4}
\begin{figure}[htbp]
\begin{center}
\centerline{\includegraphics[width=0.95\columnwidth]{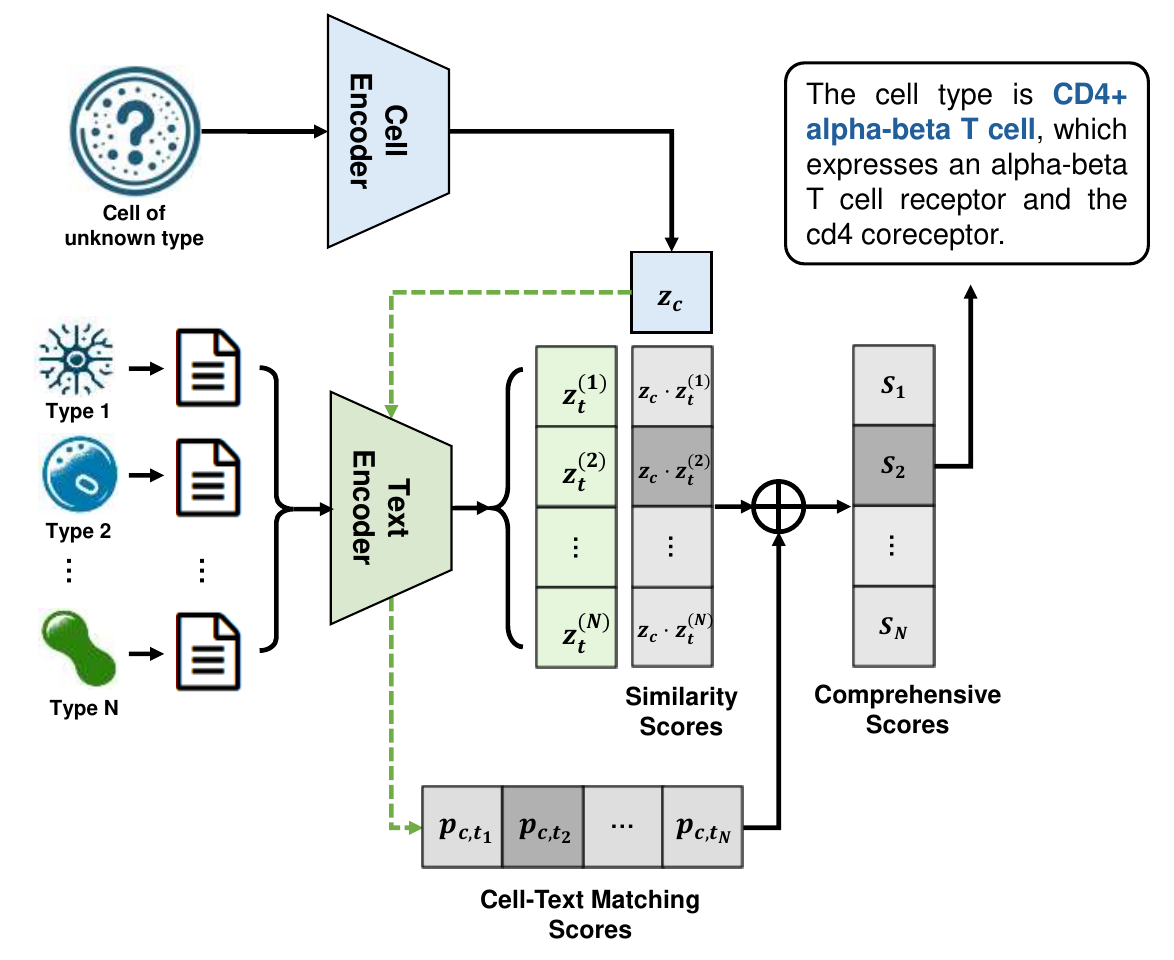}}
\caption{\textbf{The application of LangCell in zero-shot cell identity understanding.} LangCell obtains Similarity Scores using the shared embedding space of cell and text data, obtains Cell-Text Matching Scores through the matching module, and considers these comprehensively to obtain the final classification logits. In the figure, the symbol $\oplus$ represents the weighted sum after the Softmax operation.}
\label{fig2}
\end{center}
\vskip -0.1in
\end{figure}

Based on the aligned representation space of cell data and text, and utilizing the cell-text matching module with cross-attention, LangCell can be used for zero-shot cell identity understanding (Fig. \ref{fig2}). Specifically, for a given single cell \(c\) and $N$ candidate text descriptions \(\{t^{(i)}\}_{i=1}^{N}\), we obtain \(\mathrm{logits_1}\) by comparing their cosine distances in the shared embedding space, and obtain \(\mathrm{logits_2}\) by comparing the scores given by the cell-text matching module. Both are considered comprehensively for classification according to the weight of $\alpha$. In practical applications, since \(\mathrm{logits_2}\) is slower to compute, we can calculate \(\mathrm{logits_2}\) only for the candidates with high \(\mathrm{logits_1}\) scores. 

The specific calculations are as follows:
\begin{align*}
\mathrm{logits} = \alpha & \cdot \text{softmax}(\{\bm{z_c} \cdot \bm{z_t^{(i)}}\}_{i=1}^{N}) + \\ (1-\alpha) & \cdot \text{softmax}(\{\bm{g_2}(\bm{z_c}, t^{(i)})\}_{i=1}^{N})
\end{align*}
Additionally, a classification or regression head can also be added after the cell encoder for fine-tuning in downstream tasks. This downstream setting is referred to as \textbf{LangCell-CE} (\textbf{C}ell \textbf{E}ncoder) in the subsequent experiments.

\section{Experiments}

\subsection{Experiment Settings}
\subsubsection{Pre-training Details}
\textbf{Dataset Construstion}: 
We established scLibrary, a comprehensive dataset comprising roughly 27.5 million pairs of scRNA-seq data and associated textual descriptions. This dataset was sourced from the CELLxGENE database, where we acquired scRNA-seq data in raw count matrix format and the corresponding meta data. Our selection criteria included all human cells processed using the 10X sequencing protocol. We excluded data that were duplicates, had fewer than 200 expressed genes, had excessive missing metadata, or were used in downstream tasks. scRNA-seq data were processed according to \ref{sec3.1}.
Next, we selected 8 cell identities from the meta data that might contain important insights and used these entries to generate multi-view textual descriptions of cells from OBO Foundry. Specifically, the selected cell identities include assay, cell type, developmental stage, tissue information, organ information, disease information, as well as the donor's gender and ethnicity, with each entry's type detailed in the Appendix \ref{sec:app_Datasets}.

\textbf{Pre-training Stages}: Our pre-training consists of two stages. In the first stage, we initialize the cell encoder parameters using Geneformer and conduct unimodal training using only the $\mathcal{L}_{MGM}$ and $\mathcal{L}_{C-C}$ loss functions. This is to obtain a better single-cell representation learning model. In the second stage, we initialize the text encoder parameters using PubMedBERT and engage in multimodal training using all four loss functions. Both stages are trained separately for three epochs each.

\textbf{Setups}: 
The training process was conducted using the Pytorch framework and the Hugging Face transformers library. We employed the AdamW optimizer, with the learning rate warmed up to 1e-5 over 1000 steps, followed by a linear decay strategy. Weight decay was set to 0.001. More detailed parameter settings can be found in the Appendix \ref{sec:config}.

\subsubsection{Downstream Task Datasets}
We collected a set of benchmark datasets to evaluate our model's performance on different downstream tasks. These include human peripheral blood cell datasets \cite{pbmc10k, zheng2017massively},  human liver datasets \cite{scClassify}, a human brain cell dataset \cite{PCdataset}, and a comprehensive human cell dataset \cite{Tabula_Sapiens}. In addition, we propose two novel cell identity understanding tasks: non-small cell lung cancer (NSCLC) subtype classification and cell pathway identification, for which we have constructed high-quality benchmarks. The original data used to build these benchmarks were obtained from CELLxGENE. The former uses disease information from clinical diagnosis in the metadata to annotate two subtypes of NSCLC; the latter employs the irGSEA package to label hallmark pathways from the MSigDB \cite{liberzon2011molecular}. More details can be found in the Appendix \ref{sec:app_dtdata}.

\begin{table*}[ht]
\caption{\textbf{Results of zero- and few-shot cell type annotation.} LangCell is the only single-cell PLM that can perform zero-shot. All other models need to add classification headers and fine-tune. In most cases, LangCell's zero-shot performance is better than the few-shot results of existing models. Acc: accuracy (\%). $F_1$: macro $F_1$ score (\%).}
\label{nshot_table}
\begin{center}
\begin{small}
\resizebox{\textwidth}{!}{
\begin{tabular}{cl|cc|cc|cc|cc|cc|cc}
\toprule
\multirow{2}{*}{\textbf{Dataset}} & \multirow{2}{*}{\textbf{Model}} & \multicolumn{2}{c}{0-shot} & \multicolumn{2}{c}{1-shot} & \multicolumn{2}{c}{3-shot} & \multicolumn{2}{c}{5-shot} & \multicolumn{2}{c}{7-shot} & \multicolumn{2}{c}{9-shot} \\
 & & Acc & $F_1$ & Acc & $F_1$ & Acc & $F_1$ & Acc & $F_1$ & Acc & $F_1$ & Acc & $F_1$ \\

\midrule
\multirow{5}{*}{\makecell{\textbf{PBMC} \\ \textbf{10k}}}
& scBERT     & \ding{55} & \ding{55} & 31.4 & 9.4 & 60.6 & 41.4 & 78.0 & 58.2 & 59.0 & 54.6 & 81.9 & 62.6 \\
& scGPT        & \ding{55} & \ding{55} & 41.9 & 34.0 & 43.3 & 41.1 & 81.1 & 66.3 & 82.0 & 68.3 & 86.7 & 75.8 \\
& Geneformer    & \ding{55} & \ding{55} & 54.0 & 42.2 & 70.3 & 46.7 & 81.0 & 63.9 & 80.9 & 71.2 & 88.0 & 78.6\\
\cmidrule{2-14}
& \textbf{LangCell-CE} & \ding{55} & \ding{55} & \textbf{88.7} & 75.2 & 92.2 & 86.1 & 93.0 & 88.7 & 93.6 & 89.1 & 94.4 & 90.7 \\
& \textbf{LangCell} & \textbf{86.5} & \textbf{89.6} & 88.1 & \textbf{87.5} & \textbf{95.1} & \textbf{94.7} & \textbf{96.0} & \textbf{94.8} & \textbf{96.3} & \textbf{95.3} & \textbf{96.8} & \textbf{95.2} \\
\midrule  
\specialrule{0em}{0.5pt}{1pt}
\midrule

\multirow{5}{*}{\makecell{\textbf{PBMC} \\ \textbf{3\&68k}}} 
& scBERT        & \ding{55} & \ding{55} & 19.9 & 13.8 & 36.5 & 39.4 & 48.5 & 43.0 & 48.3 & 40.9 & 47.6 & 51.7\\
& scGPT         & \ding{55} & \ding{55} & 17.7 & 21.1 & 45.3 & 44.3 & 52.0 & 61.6 & 79.9 & 76.9 & 85.7 & 76.9\\
& Geneformer    & \ding{55} & \ding{55} & 21.1 & 24.7 & 55.2 & 49.2 & 59.3 & 69.1 & 81.5 & 74.8 & 83.3 & 74.1 \\
\cmidrule{2-14}
& \textbf{LangCell-CE}    & \ding{55} & \ding{55} & 86.2 & 79.7 & 85.8 & 85.8 & 87.2 & 84.1 & 89.1 & 85.9 & 86.7 & 86.0 \\
& \textbf{LangCell } & \textbf{83.9} & \textbf{82.6} & \textbf{89.7} & \textbf{87.1} & \textbf{89.9} & \textbf{87.8} & \textbf{90.3} & \textbf{87.7} & \textbf{92.1} & \textbf{87.5} & \textbf{92.4} & \textbf{88.5}  \\
\bottomrule
\end{tabular}}
\end{small}
\end{center}
\end{table*}

\subsubsection{Baselines}
Our focus is on comparison with single-cell pre-trained language models such as Geneformer, scGPT, xTrimeGene, and scBERT, which have already been proven to outperform traditional methods in a multitude of tasks. Additionally, although BioTranslator did not undergo pre-training for single-cell representation learning, it is an important point of comparison as the first model to consider leveraging textual descriptions to address single-cell issues.

\begin{figure*}
    \centering
    \includegraphics[width=\linewidth]{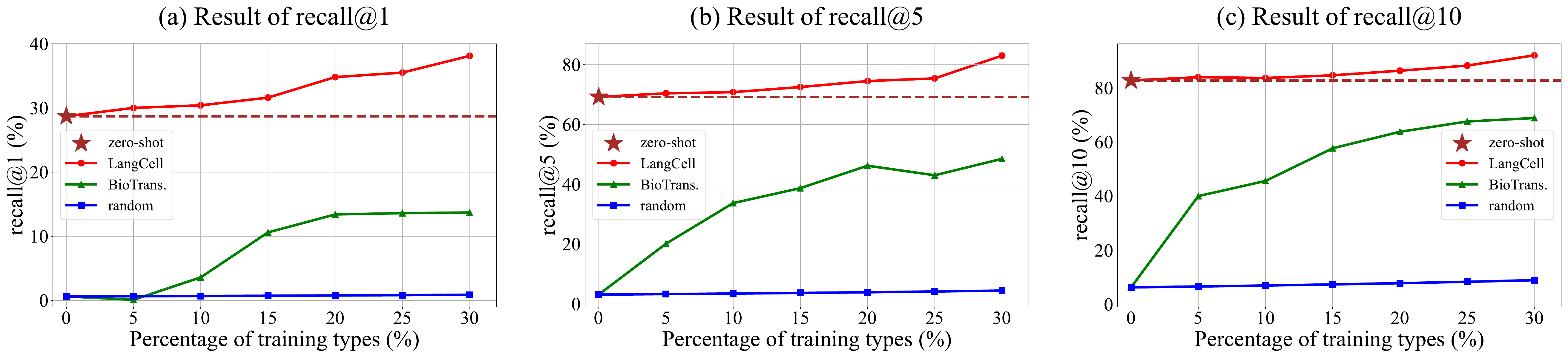}
    \caption{\textbf{Results of cell-text retrieval.} Zero-shot LangCell surpasses BioTranslator trained on up to 30\% of the 161 types.}
    \label{fig:retrieval}
\end{figure*}

\subsection{Zero-shot Cell Identity Understanding}

\subsubsection{Novel Cell Type Identification}
\label{sec:4.2.1}

\textbf{Experimental Setup}: Accurate cell type annotation is fundamental for extensive scRNA-seq analyses. However, in practical scenarios, it is often difficult to find enough high-quality labeled data for each cell type to be annotated for fine-tuning. This poses a major challenge to the application of existing single-cell models in actual scenarios, where existing models can only assign all unseen new classes under the ``\textit{Novel}'' label \cite{ScBERT_yang_2022}.
We refer to this highly challenging zero-shot task as  ``\textit{Novel Cell Type Identification}'', which requires the model to perform cell type annotation in the absence of fine-tuning data.
In addition to zero-shot learning, we engage in few-shot learning to benchmark against baselines and evaluate the data efficiency of LangCell. We meticulously selected few-shot settings that are relevant and have practical implications in bioinformatics. This study focuses on two common scenarios encountered in real-world applications:

\textit{Zero-and Few-shot Cell Type Annotation}: Suitable for scenarios with fewer alternative cell types, commonly seen in annotating small-scale single-cell data from specific tissues or dividing cells of a certain cell type into multiple subtypes. 
The few-shot approach is configured to use $n$ ($1\leq n \leq 9$) training samples for each category during fine-tuning. This configuration is designed with the practical consideration that a smaller number of alternative types in real-world settings enables the feasibility of providing very little manually annotated data for each category. 
In the few-shot task, all baseline models add a linear layer as the classification head. LangCell uses two settings: fine-tuning with the C-T and CTM tasks or using only its cell encoder and a linear classification head (\textbf{LangCell-CE}).
Our analysis is conducted on two benchmarks of human peripheral blood mononuclear cells, PBMC10K and PBMC3\&68K. The evaluation metrics employed are accuracy and macro $F_1$ score.

\textit{Cell-Text Retrieval} Suitable for scenarios with many alternative cell types, commonly seen in annotating single cells in complex environments or completely unknown single cells. 
Given the abundance of cell types, which may encompass subtypes with ambiguous boundaries, we define this task as a cell-text retrieval task. The few-shot method is structured to perform fine-tuning on a limited selection of cell types, followed by testing on a range of unseen cell types. This approach is compared with BioTranslator, the sole existing model with cross-type transfer capabilities. This setting mirrors a real-life situation where some known annotated data is available, but the target data differs from the known dataset. 
We test on the challenging human comprehensive cell dataset, Tabula Sapiens, which includes 161 cell types, 66 of which are completely new types not included in scLibrary. We use common evaluation metrics for multimodal retrieval tasks, recall@k.

\textbf{Results}: We report the test results of n-shot cell type annotation in Table \ref{nshot_table} and Fig. \ref{fig0}, and the results of cell-text retrieval in Fig. \ref{fig:retrieval}. The experimental results show that LangCell performs excellently in both zero-shot and few-shot settings for both types of tasks. Specifically, we observe:

\textit{Zero- and Few-shot Cell Type Annotation}: LangCell shows excellent performance in zero-shot scenarios, surpassing few-shot baselines in most cases. LangCell can also use a few examples to adapt to new tasks quickly, demonstrating its high data efficiency. 

\textit{Cell-Text Retrieval}: The zero-shot performance of LangCell surpasses BioTranslator, which at most uses 48 (30\% of 161) cell types, for training. This confirms LangCell's good performance in such challenging application scenarios. 

\subsubsection{NSCLC Subtype Classification}
\label{sec:4.2.2}
ScRNA-seq technology plays a significant role in the study of malignant tumor. However, it is difficult to analyze malignant cells due to the scarcity of data and their characteristics of high mutational burdens. Lung squamous cell carcinoma (LUSC) and lung adenocarcinoma (LUAD) are the two most common subtypes of non-small cell lung cancer (NSCLC). We test LangCell on 2,658 malignant cells from patients with these two lung cancer subtypes, to assess its effectiveness in identifying disease-related cell identities.

\begin{figure}[t]
\begin{center}
\centerline{\includegraphics[width=0.85\columnwidth]{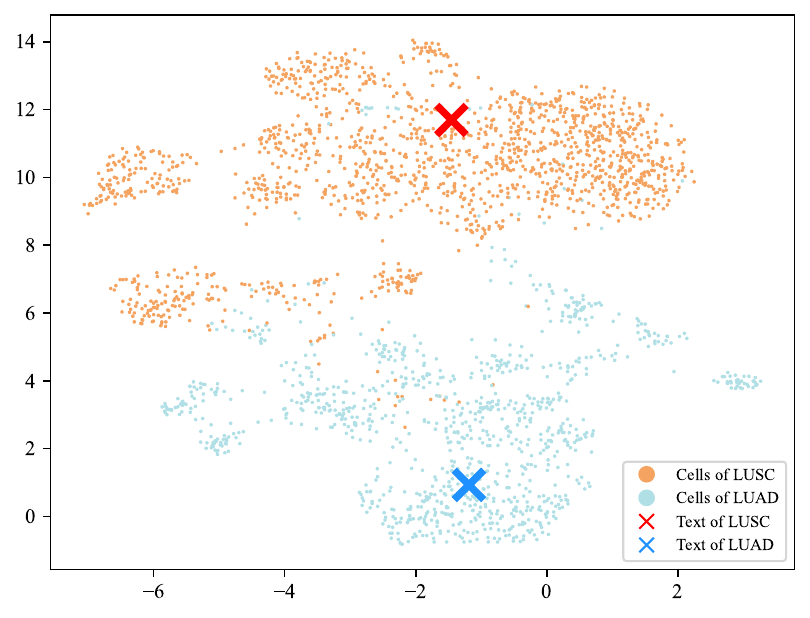}}
\caption{UMAP plot of embeddings for scRNA-seq data and descriptions of two NSCLC subtypes.}
\label{umap_cancer}
\end{center}
\vskip -0.2in
\end{figure}

Since all cell type labels are ``\textit{Malignant}'', we used descriptions of these two diseases to construct the texts. As shown in Figure \ref{umap_cancer}, LangCell aligns single-cell and disease texts well in the latent space. 
Its zero-shot classification surpasses Geneformer (fine-tuned with 10-shot learning) by about 20\% in both accuracy and macro $F_1$-score. (Table \ref{cancer_classifacation_table})
This experiment demonstrates LangCell's strong capability in understanding disease-related cell identities and its effective performance in the analysis of single cells with high mutational burdens, such as malignant cells.

\begin{table}[t]
\begin{center}
\begin{small}
\caption{Results of NSCLC subtype classification (\%).}
\label{cancer_classifacation_table}
\resizebox{0.3\textwidth}{!}{
\begin{tabular}{ll|cc}
        \toprule
        \textbf{Model} & n-shot & Acc & $F_1$\\
        \midrule
        \multirow{2}{*}{\textbf{Geneformer}}
        & 1 & 46.7 & 43.9 \\
        & 10 & 73.1 & 73.1 \\
        \midrule
        \textbf{LangCell} & 0 & \textbf{93.5} & \textbf{93.2} \\
        \bottomrule
    \end{tabular}}
\end{small}
\end{center}
\vskip -0.05in
\end{table}

\subsubsection{Single-cell batch integration} 
Single-cell batch integration holds significant importance in biomedical research. It plays a crucial role in mitigating batch effects from different experimental data, scaling up data analysis, and fostering a comprehensive understanding of cell diversity and functionality within biological systems. This task requires the model to correctly distinguish whether the expression differences between cells arise from meaningless batch effects or meaningful biological information, demanding a strong understanding of cell identity information inherent in scRNA-seq data.

We evaluated the performance of LangCell in cell batch integration on the PBMC10K and Perirhinal Cortex, comparing it with the classical model in the field, scVI, as well as several single-cell PLMs. To comprehensively assess model performance, we used the evaluation metrics Avg$_{bio}$, ASW$_{batch}$, and S$_{final}$ proposed in \cite{scib}. These metrics respectively assess the model's capability of biological integration, batch effect removal, and a comprehensive evaluation of both. Detailed calculation can be found in the Appendix \ref{sec:metrics}.
The experimental results in Table \ref{cluster_table} indicate that LangCell surpasses the existing optimal models in all three metrics. This demonstrates LangCell's profound insight into transcriptomic data, its ability to accurately preserve important biological information, and its effectiveness in correcting irrelevant batch effects.

\begin{table}[t]
\caption{Results of cell batch integration (\%) from scratch. }
\label{cluster_table}
\begin{center}
\begin{small}
\resizebox{0.4\textwidth}{!}{
\begin{tabular}{clccc}
\toprule
\textbf{Dataset} & \textbf{Model} & Avg$_{bio}$ & ASW$_{batch}$ & S$_{final}$ \\
\midrule
\multirow{5}{*}{\textbf{PBMC10K}}
& scVI          & 70.0 & 97.6 & 81.0  \\
& scBERT        & 18.1 & 95.0 & 48.9 \\
& scGPT*       & 72.3 & 91.9 & 80.2 \\
& Geneformer    & 79.3 & 92.8 & 84.7 \\
& \textbf{LangCell} & \textbf{80.8} & \textbf{97.9} & \textbf{87.6} \\
\midrule  
\specialrule{0em}{0.5pt}{1pt}
\midrule
\multirow{5}{*}{\makecell{\textbf{Perirhinal} \\ \textbf{Cortex}}} 
& scVI          & 84.9 & 89.6 & 86.8 \\
& scBERT        & 15.1 & 92.9 & 46.2 \\
& scGPT*          & 88.9 & 88.4 & 88.7 \\
& Geneformer    & 85.5 & 91.8 & 88.0 \\
& \textbf{LangCell} & \textbf{95.2} & \textbf{95.6} & \textbf{95.4}\\

\bottomrule
\end{tabular}}
\end{small}
\end{center}
\end{table}

\begin{table}[t]
\caption{Results of cell type annotation (\%). *: Since xTrimoGene did not release the checkpoint, we can only obtain their reported Zheng68k result.}
\label{cell_type_annotation_table}
\begin{center}
\begin{small}
\resizebox{0.5\textwidth}{!}{
\begin{tabular}{l|cccccc}
\toprule
\multirow{2}{*}{\textbf{Dataset}} & \multicolumn{2}{c}{PBMC10K} & \multicolumn{2}{c}{LiverCross} & \multicolumn{2}{c}{Zheng68K} \\
 & Acc & $F_1$ & Acc & $F_1$ & Acc & $F_1$ \\
\midrule  
\specialrule{0em}{0.5pt}{1pt}
\midrule
scBERT       & 97.5 & 90.5 & 37.3 & 12.2 & 77.9 & 68.8   \\
scGPT        & 96.5 & 94.1 & 48.1 & 24.1 & 84.6 & 75.2  \\
xTrimoGene* & - & - & -  & -  & -  & 73.5   \\
Geneformer   & 97.8 & 95.7 & 46.7 & 24.0 & 83.9 & 74.4 \\
\textbf{LangCell-CE} & \textbf{98.3} & \textbf{96.9} & \textbf{50.4} & \textbf{26.0} & \textbf{85.4} & \textbf{76.9} \\
\bottomrule
\end{tabular}}
\end{small}
\end{center}
\end{table} 

\subsection{Cell Representation Learning of LangCell-CE}

\subsubsection{Cell Type Annotation (fine-tune)} 

We also evaluate the representation capabilities of LangCell-CE on the classic task of cell type annotation. The results in Table \ref{cell_type_annotation_table} demonstrate that our model achieves SOTA performance on all three datasets. 
This demonstrates that LangCell successfully injects unstructured knowledge into the cell encoder, enhancing the understanding of scRNA-seq data.
In addition, experiments on LiverCross demonstrate the effectiveness of LangCell in cross-dataset tasks.

\subsubsection{Pathway Identification}
In the pre-training process, we injected knowledge of ``cell types'' into LangCell-CE, thereby naturally deepening its understanding of this cell identity. We cannot yet determine whether LangCell-CE's outstanding performance is due to its comprehensively improved ability to learn cell representations, or simply due to its insights into this specific task.  
To verify this, we explored a new cell identity not covered in pre-training, cell pathways, and designed a challenging representation learning task around it. For each single cell, the model was tasked with identifying multiple pathways from a selection of 41 important pathways, essentially constituting a multi-label binary classification problem with 41 independent labels. Considering the data imbalance caused by the expression of each pathway in only a few cells, focal loss \cite{lin2017focal} was used during fine-tuning. As shown in Table \ref{pathway}, LangCell significantly outperforms the existing state-of-the-art model Geneformer on AUROC and AUPRC on this challenging task. 

\begin{table}[t]
\begin{center}
\begin{small}
\caption{Result of pathway identification (\%). The metrics are detailed in Appendix \ref{pathway}.}
\label{pathway}
\resizebox{0.48\textwidth}{!}{
\begin{tabular}{lcccc}
        \toprule
        \textbf{Model} & avg-AUROC & avg-AUPRC & flatten-AUROC & flatten-AUPRC \\
        \midrule
        Geneformer & 82.8 & 23.9 & 86.6 & 27.3\\
        \midrule
        \textbf{LangCell-CE} & \textbf{89.3} & \textbf{31.4} & \textbf{89.9} & \textbf{35.4} \\
        \bottomrule
    \end{tabular}}
\end{small}
\end{center}
\end{table}

\begin{table}[t]
\centering
\small
\caption{Ablation study of pre-training tasks in LangCell. \textit{LangCell-1}: model at the end of the first stage of pre-training. \textit{w/o CTM}: without  CTM module.}
\resizebox{0.5\textwidth}{!}{
    \begin{tabular}{l|cccc|cc|cc}
    \toprule
    \multirow{2}{*}{\textbf{Models}} & \multirow{2}{*}{$\mathcal{L}_{MGM}$} & \multirow{2}{*}{$\mathcal{L}_{C-C}$} & \multirow{2}{*}{$\mathcal{L}_{C-T}$} & \multirow{2}{*}{$\mathcal{L}_{CTM}$} & \multicolumn{2}{c|}{zero-shot} & \multicolumn{2}{c}{fine-tune} \\
      &&&&& Acc & $F_1$ & Acc & $F_1$ \\
     \midrule
    Geneformer & \color{teal}\ding{52} & \color{red}\ding{55} & \color{red}\ding{55} & \color{red}\ding{55} & - & - & 76.1 & 64.7 \\
    LangCell-1 & \color{teal}\ding{52} & \color{teal}\ding{52} & \color{red}\ding{55} & \color{red}\ding{55} & - & - & 77.0 & 65.8 \\
    LangCell $_{\mathrm{w/o\,CTM}}$ & \color{teal}\ding{52} & \color{teal}\ding{52} & \color{teal}\ding{52} & \color{red}\ding{55} & 84.8 & 85.9 & - & - \\
    \textbf{LangCell} & \color{teal}\ding{52} & \color{teal}\ding{52} & \color{teal}\ding{52} & \color{teal}\ding{52} & \textbf{85.2} & \textbf{86.1} & \textbf{78.0} & \textbf{66.6} \\ 
    \bottomrule
    \end{tabular}}
\label{tab:ablation}
\end{table}

\subsection{Ablation Study}
\begin{figure}[t]
\begin{center}
\centerline{\includegraphics[width=0.8\columnwidth]{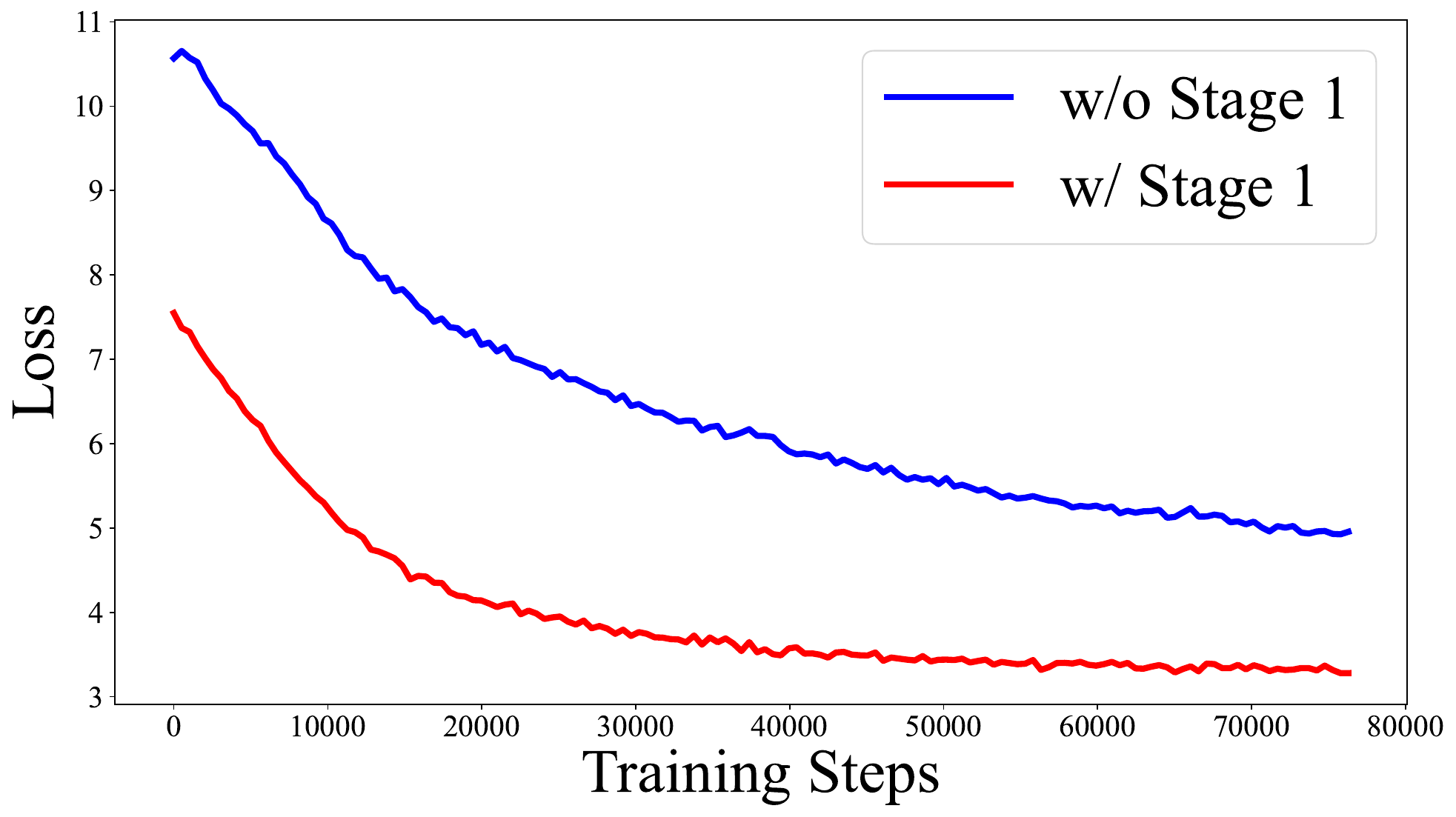}}
\caption{The impact of whether to conduct the first stage on the convergence speed of LangCell pre-training.}
\label{loss}
\end{center}
\vskip -0.1in
\end{figure}

\paragraph{Stage Division:} The primary motivation for dividing the training into two stages is to reduce computational costs. Our preliminary experiments showed that the first stage of training significantly accelerates the convergence of the loss function in the second phase (Fig. \ref{loss}), thereby reducing the number of epochs needed to reach convergence. Considering that the first phase only requires computing $\mathcal{L}_{MGM}$ and $\mathcal{L}_{C-C}$, whose computational costs are much lower than $\mathcal{L}_{C-T}$ and $\mathcal{L}_{CTM}$, the two-stage pre-training significantly reduces the overall computational resources needed.

\paragraph{Initialization:} If we don't use Geneformer and PubMedBERT to initialize the model parameters and instead start training from scratch, the convergence of the model will be extremely slow in the early stages due to the absence of high-quality representations of scRNA-seq data and text. Specifically, the loss demonstrates negligible decline throughout the first 20,000 steps, and the huge computational costs prevent further study of training from scratch.

\paragraph{Ablation Study of Pre-training Tasks:} We explored the influence of various pre-training tasks on the model's performance in downstream applications (Table \ref{tab:ablation}). Compared to using only $\mathcal{L}_{MGM}$, the incorporation of $\mathcal{L}_{C-C}$ improves the model's cell representation learning capability, which is consistent with findings in NLP \cite{gao-simcse} and CV \cite{what_do_sfvt} fields. The integration of textual data further enhances the performance of the cellular encoder, which we attribute to the injection of unstructured knowledge in the text.
The results on zero-shot and few-shot retrieval tasks demonstrate the importance of considering both similarity scores and matching scores comprehensively, with performance surpassing that of considering either alone.

\paragraph{Discussion on the setting of $\alpha$:} $\alpha$ mentioned in \ref{sec3.4} is an adjustable hyperparameter during downstream tasks. Users of LangCell can use a small validation set to select the optimal $\alpha$ for a specific task. If there is no labeled validation set available, we recommend setting 0.2 as the default value for $\alpha$, which is near-optimal in most cases. In the zero-shot experiments of this paper, to simulate an application scenario with no labeled data at all, we did not manually adjust $\alpha$ and instead used the default value 0.2. The results in Table \ref{tab:alpha} present the impact of $ \alpha$ on several zero-shot tasks.

\begin{table}[t]
\centering
\small
\caption{The impact of alpha setting on the model's zero-shot ability.}
\resizebox{0.4\textwidth}{!}{
\begin{tabular}{lcccc}
\toprule
\multicolumn{1}{c}{\textbf{}} & \multicolumn{2}{c}{\textbf{PBMC10K}} & \multicolumn{2}{c}{\textbf{PBMC3\&68K}} \\
$\alpha$ & Accuracy & $F_1$ & Accuracy & $F_1$ \\
\midrule
0 & 56.83 & 25.09 & 65.32 & 31.33 \\
0.01 & 86.74 & 67.07 & \textbf{88.54} & 81.45 \\
0.05 & \textbf{92.40} & 80.08 & 87.44 & \textbf{83.07} \\
0.1 & 90.98 & 85.18 & 85.13 & 82.47 \\
0.2 & 86.54 & \textbf{89.61} & 83.94 & 82.64 \\
0.3 & 86.29 & 89.47 & 84.25 & 82.16 \\
0.5 & 85.98 & 89.49 & 83.91 & 82.39 \\
0.7 & 85.84 & 89.46 & 83.74 & 82.42 \\
0.9 & 85.77 & 89.43 & 83.70 & 82.43 \\
\bottomrule
\end{tabular}}
\label{tab:alpha}
\vskip 0.15in
\end{table}

\section{Conclusion and Limitation}

In this work, we present LangCell, the Language-Cell pre-training framework, offering a unified representation of single-cell data and natural language that transcends the need for task-specific fine-tuning. By integrating these modalities, LangCell intuitively grasps the relationship between cellular data and textual identities, enhancing its cell representation learning capabilities. Our experiments across various biological tasks confirm LangCell's superior performance over existing models, particularly in zero-shot and few-shot scenarios. This framework sets a new standard for the field, enabling more accurate and efficient analysis of single-cell transcriptomics data.

Currently, LangCell still has some limitations. For example, the pre-training texts are all from the OBO Foundry, which limits the diversity to a certain extent. LangCell also cannot yet analyze other single-cell omics such as scATAC-seq, and it does not include cell/text generation functions. In the future, we will focus on improving these aspects.

\section*{Code Availability}
LangCell will soon be added to the OpenBioMed toolkit: \href{https://github.com/PharMolix/OpenBioMed/}{https://github.com/PharMolix/OpenBioMed}.

Code is available at: \href{https://github.com/PharMolix/LangCell}{https://github.com/PharMolix/LangCell}.

\section*{Acknowledgements}
This research is supported by the National Key R\&D Program of China (No. 2022YFF1203002) and PharMolix Inc.

\section*{Impact Statement}
This paper presents work whose goal is to advance the field of Machine Learning. There are many potential societal consequences of our work, none of which we feel must be specifically highlighted here.


\balance
\clearpage\clearpage


\bibliography{icml2023_ref}

\bibliographystyle{icml2024}
\balance

\newpage
\appendix
\onecolumn
\setcounter{section}{0}
\setcounter{equation}{0}
\setcounter{subsection}{0}
\renewcommand{\theequation}{A.\arabic{equation}}
\renewcommand{\thetable}{\Alph{table}}
\renewcommand{\thefigure}{\Alph{figure}}
\numberwithin{figure}{subsection}
\numberwithin{table}{subsection}
\section*{Appendix}

\section{More Experimental Results}

\subsection{Cell Batch Integration \& Novel Cell Type Identification}
The complete experimental results for cell batch integration are shown in Table \ref{cluster_table_full}.

\begin{table}[ht]
\caption{Results of cell batch integration (\%) from scratch. * stands for the results from scGPT.}
\label{cluster_table_full}
\vskip 0.15in
\begin{center}
\begin{small}
\resizebox{0.8\textwidth}{!}{
\begin{tabular}{clcccccc}
\toprule
\textbf{Dataset} & \textbf{Model} & NMI & ARI & ASW$_{cell}$ & Avg$_{bio}$ & ASW$_{batch}$ & S$_{final}$ \\
\midrule
\multirow{5}{*}{\textbf{PBMC10K}}
& scVI          & 80.8 & 71.1 & 58.1 & 70.0 & 97.6 & 81.0  \\
& scBERT        & 5.3 & 3.4 & 45.5 & 18.1 & 95.0 & 48.9 \\
& scGPT*       & 73.8 & 79.3 & 63.9 & 72.3 & 91.9 & 80.2 \\
& Geneformer   & 82.5 & 84.6 & 70.9 & 79.3 & 92.8 & 84.7 \\
& \textbf{LangCell} & \textbf{84.5} & \textbf{85.4} & \textbf{72.4} & \textbf{80.8} & \textbf{97.9} & \textbf{87.6} \\
\midrule  
\specialrule{0em}{0.5pt}{1pt}
\midrule
\multirow{5}{*}{\makecell{\textbf{Perirhinal} \\ \textbf{Cortex}}} 
& scVI          & 95.0 & 95.7 & 63.9 & 84.9 & 89.6 & 86.8 \\
& scBERT        & 3.1 & 2.7 & 39.6 & 15.1 & 92.9 & 46.2 \\
& scGPT*          & 88.6 & 89.5 & 88.6 & 88.9 & 88.4 & 88.7 \\
& Geneformer    & 89.0 & 81.3 & 86.3 & 85.5 & 91.8 & 88.0 \\
& \textbf{LangCell} & \textbf{97.2} & \textbf{98.3} & \textbf{90.2} & \textbf{95.2} & \textbf{95.6} & \textbf{95.4}\\

\bottomrule
\end{tabular}}
\end{small}
\end{center}
\vskip -0.27in
\end{table}

\begin{figure}[htp]
    \centering
    \includegraphics[width=1\linewidth]{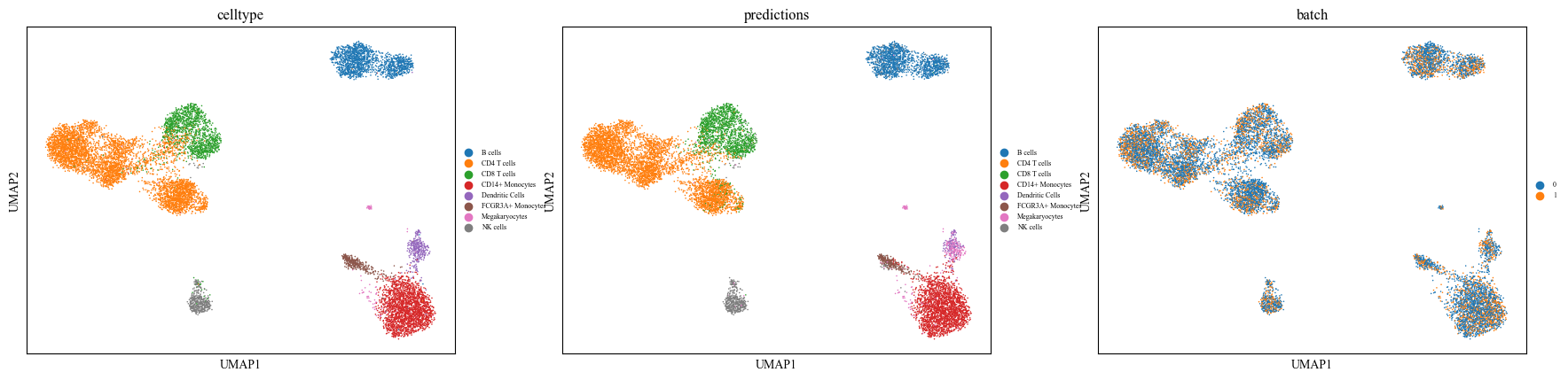}
    \caption{UMAP plot of embeddings for scRNA-seq data of LangCell in the zero-shot scenario. Three scatter plots are colored by actual cell type labels, predicted cell type labels, and batch information, respectively. It is evident that the cell embeddings generated directly by LangCell, without any fine-tuning, possess desirable properties: they cluster by cell type and eliminate batch effects. By comparing the left and middle plots, one can intuitively observe that LangCell is capable of reliably annotating cell types in a zero-shot scenario.}
    \label{umap}
\end{figure}

\begin{figure}[htp]
\begin{center}
\centerline{\includegraphics[width=1\columnwidth]{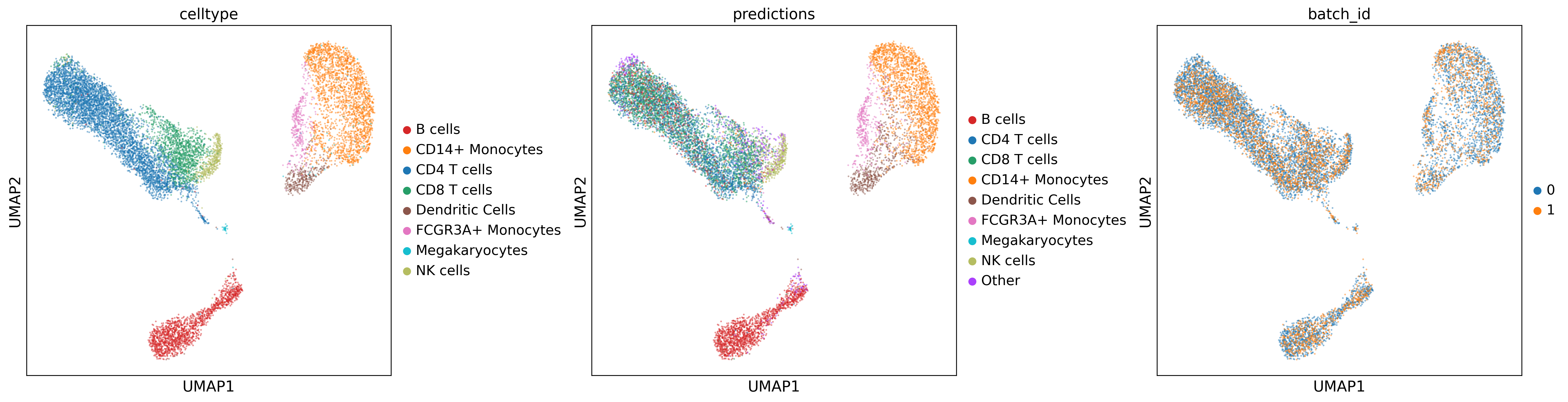}}
\caption{UMAP plot of embeddings for scRNA-seq data of scGPT. In the middle is the predicted result of fine-tuned scGPT.}
\label{umap1}
\end{center}
\end{figure}

\begin{figure}[htp]
\begin{center}
\centerline{\includegraphics[width=0.9\columnwidth]{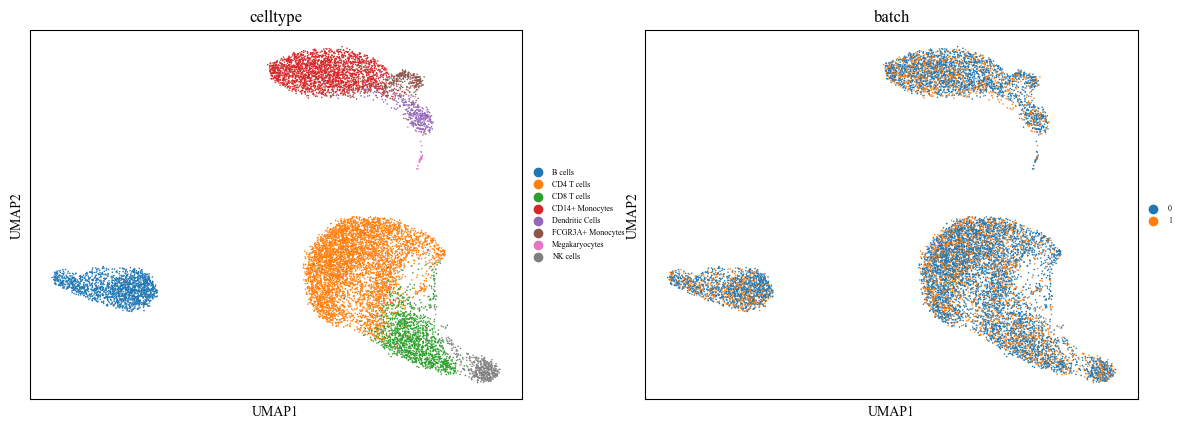}}
\caption{UMAP plot of embeddings for scRNA-seq data of Geneformer. }
\label{umap2}
\end{center}
\end{figure}

We perform a visual analysis of the PBMC10K dataset to intuitively observe LangCell's zero-shot capability for cell batch integration (Fig. \ref{umap}, left and right). We also visualize the encoding results of the current best models, scGPT and Geneformer (Fig. \ref{umap1}, Fig. \ref{umap2}). It can be observed that all three models excel at eliminating batch effects, but LangCell, during encoding, can directly focus on the identity information of cells, with cells of the same type clustering together in the feature space.

Moreover, LangCell can effectively complete novel cell type identification. Comparing the left and middle images of Fig. \ref{umap}, it can be seen that LangCell can correctly annotate most cells without any fine-tuning.

\subsection{Cell-Text Retrieval}
\begin{figure*}
\centering
\begin{minipage}{0.48\textwidth}
    \centering
    \includegraphics[width=\linewidth]{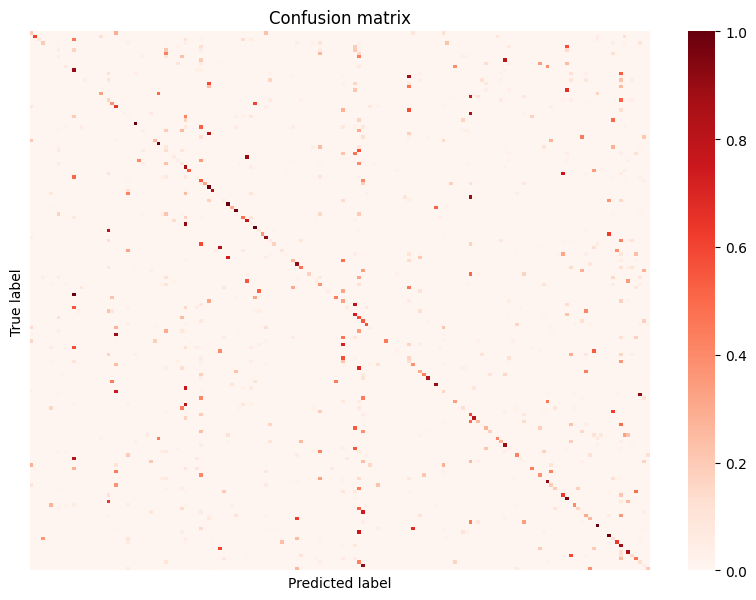} 
    \caption{Heatmap of the top-ranked results from LangCell's zero-shot retrieval on Tabula Sapiens.}
    \label{heat1}
\end{minipage}\hfill
\begin{minipage}{0.48\textwidth}
    \centering
    \includegraphics[width=\linewidth]{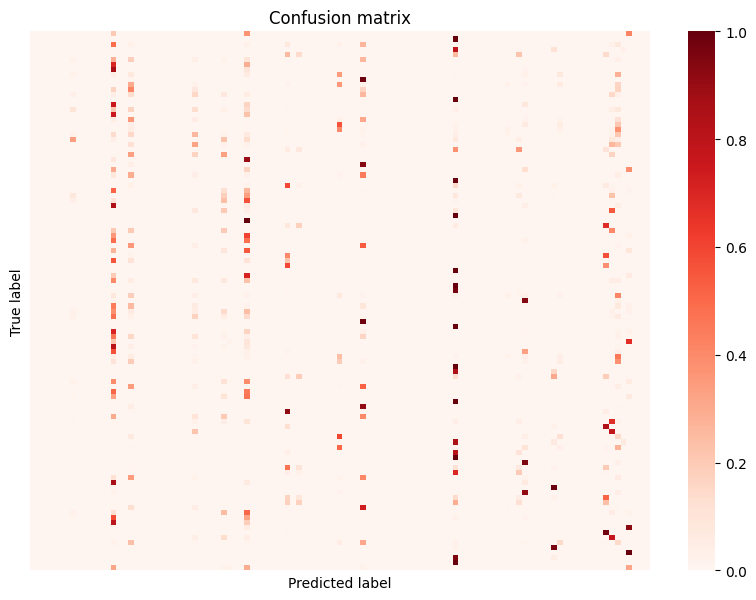} 
    \caption{Heatmap of the top-ranked results from BioTranslator's retrieval on Tabula Sapiens. Trained on 10\% types and tested on 90\% types.}
    \label{heat2}
\end{minipage}
\end{figure*}

We plotted heatmaps of top retrieval results for zero-shot LangCell and BioTranslator trained on 10\% types. It is clear to see that LangCell's result plot has a sharper diagonal line, signaling a significantly higher retrieval ability for new cell types than BioTranslator (Fig. \ref{heat1}, Fig. \ref{heat2}).

\subsection{Pathway Identification}
\begin{figure*}
\centering
\begin{minipage}{0.48\textwidth}
    \centering
    \includegraphics[width=\linewidth]{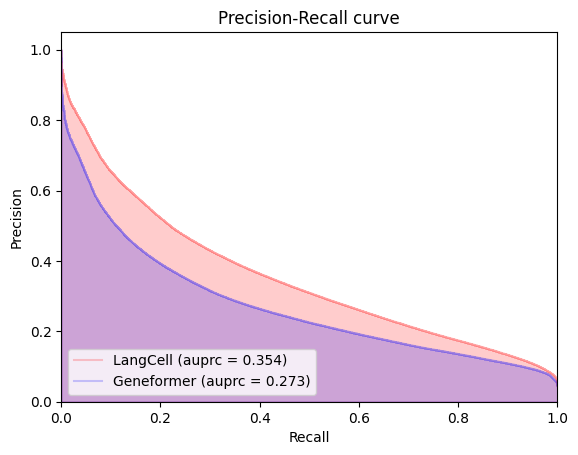}
    \caption{flatten-PRC of pathway identification.}
    \label{prc}
\end{minipage}\hfill
\begin{minipage}{0.48\textwidth}
    \centering
    \includegraphics[width=\linewidth]{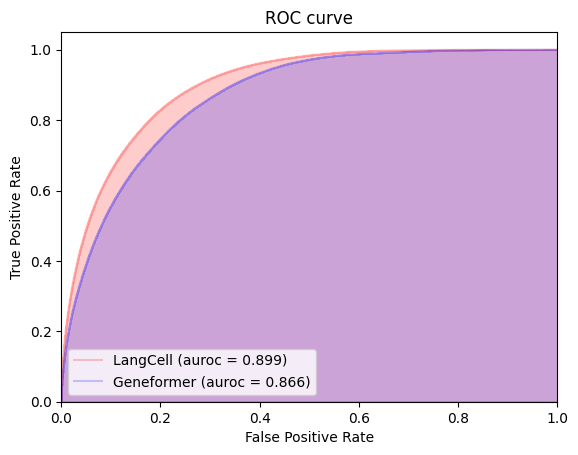}
    \caption{flatten-ROC of pathway identification.}
    \label{roc}
\end{minipage}
\end{figure*}

Fig. \ref{prc} and Fig. \ref{roc} visually show the performance of fine-tuned LangCell and Geneformer in the pathway identification task.

\subsection{Ablation Study of Pre-training Tasks}
The complete experimental results for cell batch integration are shown in Table \ref{tab:ablation_full}.

\begin{table}[t  ]
\centering
\small
\caption{Ablation study of pre-training tasks in LangCell. \textit{LangCell-1}: model at the end of the first stage of pre-training. \textit{w/o CTM}: without  CTM module.}
\resizebox{\textwidth}{!}{
    \begin{tabular}{l|cccccc|cccccccc}
    \toprule
    \multirow{4}{*}{\textbf{Models}} & \multicolumn{6}{c|}{Zero-shot} & \multicolumn{8}{c}{Fine-tune} \\
    \\ & \multicolumn{2}{c}{PBMC10K} & \multicolumn{2}{c}{PBMC3\&68K} & \multicolumn{2}{c|}{Avg} & \multicolumn{2}{c}{PBMC10K} & \multicolumn{2}{c}{LiverCross} & \multicolumn{2}{c}{Zheng68k} & \multicolumn{2}{c}{Avg} \\
      & Acc & $F_1$ & Acc & $F_1$ & Acc & $F_1$& Acc & $F_1$ & Acc & $F_1$ & Acc & $F_1$ & Acc & $F_1$ \\
     \midrule
    Geneformer  & - & - & - & - & - & - & 97.8 & 95.7 & 46.7 & 24.0 & 83.9 & 74.4 & 76.1 & 64.7 \\
    LangCell-1  & - & - & - & - & - & - & 98.1 & 96.6 & 48.5 & 25.4 & 84.4 & 75.4 & 77.0 & 65.8 \\
    LangCell $_{\mathrm{w/o\,CTM}}$  & 85.8 & 89.4 & 83.7 & 82.4 & 84.8 & 85.9  & - & - & - & - & - & - & - & - \\
    \textbf{LangCell}  & \textbf{86.5} & \textbf{89.6} & \textbf{83.9} & \textbf{82.6} & \textbf{85.2} & \textbf{86.1} & \textbf{98.3} & \textbf{96.9} & \textbf{50.4} & \textbf{26.0} & \textbf{85.4} & \textbf{76.9} & \textbf{78.0} & \textbf{66.6} \\
    \bottomrule
    \end{tabular}}
\vskip -0.1in
\label{tab:ablation_full}
\end{table}

\subsection{Retrieval for novel cell types not covered by scLibrary}
LangCell has demonstrated excellent performance in cell-text retrieval tasks. However, many cell types in the test dataset Tabula Sapiens are covered by scLibrary. To confirm that LangCell's outstanding performance is not solely due to encountering the same cell types in scLibrary, we re-calculate the experimental results in Figure \ref{fig:retrieval} for 95 cell types covered by scLibrary and 66 cell types not covered by scLibrary. We present the experimental effects of LangCell in a zero-shot scenario and compare them with the results of BioTranslator under the setting of 30\% training classes (Table \ref{tab:not_covered}). Each set of experiments uses all 161 types as alternatives. The experimental results show that for new types of cells present in Tabula Sapiens that are not included in scLibrary, LangCell also exhibits outstanding classification performance. This demonstrates LangCell's strong transferability to entirely new cell types.

\begin{table}[t]
\caption{The cell-text retrieval results of the cell types covered and not covered by scLibrary.}
\label{tab:not_covered}
\resizebox{\textwidth}{!}{
\begin{tabular}{llccc}
\toprule
\textbf{Model} & \textbf{Data Selection} & \textbf{Classes of Cells} & \textbf{Number of Cells} & \textbf{Recall@1 (Accuracy)} \\ 
\midrule
BioTranslator (baseline) & 30\% classes for training and 70\% for test & 113 & 212k & 13.71 \\
LangCell & All & 161 & 456k & 28.65 \\
LangCell & Covered by scLibrary & 95 & 429k & 28.77 \\
LangCell & Not covered by scLibrary & 66 & 27k & 26.74 \\
\bottomrule
\end{tabular}
}
\end{table}

\subsection{Robustness to ``Dropout Zeros''}
 In practical application scenarios, scRNA-seq data often contains ``dropout zeros'' noise, which means that low gene expressions may not be captured during sequencing \cite{DZ3_Silverman2018NaughtAZ}. The model's resistance to such noise significantly influences its practicality. In fact, ``dropout zeros'' can be regarded as random noise introduced by sequencing technology. Works such as Geneformer and scGPT have demonstrated that single-cell language models can understand the contextual relationships of gene expressions during large-scale pre-training, thereby possessing resistance to noise in scRNA-seq data. To verify LangCell's resistance to ``dropout zeros'', we have added the following experiment on the PBMC10K dataset. For genes with expression levels in the bottom 30\% of each cell (excluding genes with an expression level of 0), we reduced their expression levels to 0 with a certain probability to simulate the ``dropout zeros'' noise. Subsequently, we observed the perturbation of LangCell-generated cell embeddings under different probabilities of dropout zeros. We also tested the impact of different probabilities of dropout zeros on the downstream task effects in the zero-shot cell type annotation task, reflecting LangCell's performance on lower-quality downstream task data. The experiments were conducted three times for each dropout probability, and the average results are shown in Table \ref{tab:dropout_zeros}.

 The experimental results indicate that applying dropout perturbation does not cause significant shifts in the cell embeddings generated by LangCell. This demonstrates that LangCell has excellent noise resistance capabilities against the ``dropout zeros'' phenomenon specific to scRNA-seq data. Moreover, the experimental results of zero-shot cell type annotation also show that dropout zeros events in downstream task data do not significantly degrade LangCell's performance. This proves the robustness of LangCell on data of lower quality.

\begin{table}[t]
\centering
\small
\caption{The impact of ``dropout zeros'' on LangCell experimental results.}
\label{tab:dropout_zeros}
\begin{tabular}{lccc}
\toprule
\textbf{Dropout Probability (\%)} & \textbf{CosineSimilarity (\%)} & \textbf{Accuracy(\%)} & \textbf{$F_1$ (\%)} \\
\midrule
0 & 100 & 86.54 & 89.61 \\
1 & 99.70 & 86.66 & 89.59 \\
5 & 98.72 & 86.37 & 89.71 \\
10 & 97.45 & 88.14 & 89.50 \\
\bottomrule
\end{tabular}
\end{table}

\section{More Discussion about Complexity Analysis and Inference Speed}

\subsection{Complexity analysis of text encoder}

Let the text length be $N$, the gene sequence length of a cell be $M$, and consider the vector dimension as a constant. When the text encoder adopts a single-modality mode ($\bm{g_{1}}$), the bottleneck for time and space complexity lies in the self-attention computation of the text, with a complexity of $O({N}^{2})$ \cite{vaswani2023attention}. When the text encoder adopts a multi-modality mode ($\bm{g_{2}}$), the time and space complexity is determined by both the self-attention computation of the text and the cross-attention computation between the image and text, with a complexity of $O({N}^{2} + MN)$.

Furthermore, during the zero-shot inference process, let the total number of categories be $P$ and the total number of cells be $Q$. Typically, $P \ll Q$. The time complexity for calculating the embeddings of all categories in single-modality mode is $O(PN^2)$. For each cell, the time complexity for computing the embeddings is $O(M^2)$, and the time complexity for calculating the match scores with the top k categories using $\bm{g_{2}}$ is $O(k{N}^{2} + kMN)$. Therefore, the total time complexity of the inference process is $O(PN^2 + Q (M^2 + kN^2 + kMN))$.

\subsection{Discussion about inference speed}
The total time complexity of the reasoning process is $O(PN^2 + Q (M^2 + kN^2 + kMN))$. Put simply, the total number of forward passes required during the inference process is $P + Q + kQ$, where $0 \leq k \leq P$.

The number of forward passes for LangCell-CE, Geneformer and other models that require fine-tuning during inference is $Q$. Considering $P \ll Q$, the main factor affecting inference time is the choice of $k$. In scenarios where inference speed is highly required, $k$ can be set to 0, thus achieving fast inference similar to LangCell-CE; in scenarios with a larger number of categories or where inference speed is not a high requirement, a larger $k$ can be chosen or even $k$ can be set to $P$ for more accurate inference.

For the experimental results reported in the paper, except for the cell-text retrieval experiment where $k=20$ is taken, all others take $k=P$. In the ablation study reported in Table \ref{tab:ablation}, LangCell $_{\bm{\mathrm{w/o CTM}}}$ reports the average zero-shot performance of the model on the two datasets in Table \ref{nshot_table} under the setting of $k=0$. In Table \ref{tab:k_value_impact}, we provide the impact of different $k$ values on the model's inference performance and time on the PBMC10K dataset ($P=8$).

\begin{table}[t]
\centering
\small
\caption{The impact of $k$ on the model's inference performance and time. ($P=8$)}
\label{tab:k_value_impact}
\begin{tabular}{cccc}
\toprule
\textbf{k} & \textbf{Time (s)} & \textbf{Zero-Shot Accuracy (\%)} & \textbf{Zero-Shot $F_1$ (\%)} \\
\midrule
0 & \textbf{186.62} & 86.76 & 89.37 \\
2 & 514.51 & 86.28 & 89.23 \\
4 & 841.18 & 86.04 & 89.49 \\
6 & 1162.49 & 86.53 & 89.55 \\
8 & 1537.91 & \textbf{86.54} & \textbf{89.61} \\
\midrule
\specialrule{0em}{0.5pt}{1pt}
\midrule
\multicolumn{4}{l}{For comparison, the inference time of LangCell-CE or Geneformer: 188.23s.}\\
\bottomrule
\end{tabular}
\end{table}

Experimental results are consistent with theoretical derivations, demonstrating that larger values of $k$ enhance model performance but also increase the time cost of inference. Furthermore, the inference speed at $k=0$ is close to that of LangCell-CE or Geneformer. Fortunately, in most cases, the model performance at $k=0$ is not much lower than at $k=P$, and it generally still surpasses other methods. When users have high demands for inference speed, setting $k=0$ is a viable option to quickly obtain satisfactory results. In Table \ref{nshot_complete}, we provide the complete performance of LangCell at $k=0$ (i.e., w/o CTM) in the experiments of Table \ref{nshot_table} from the paper.

\begin{table}[t]
\caption{The complete performance of LangCell at $k=0$ (i.e., w/o CTM).}
\label{nshot_complete}
\small
\centering
\resizebox{0.9\textwidth}{!}{
\begin{tabular}{cl|cc|cc|cc|cc|cc|cc}
\toprule
\multirow{2}{*}{\textbf{Dataset}} & \multirow{2}{*}{\textbf{Model}} & \multicolumn{2}{c}{0-shot} & \multicolumn{2}{c}{1-shot} & \multicolumn{2}{c}{3-shot} & \multicolumn{2}{c}{5-shot} & \multicolumn{2}{c}{7-shot} & \multicolumn{2}{c}{9-shot} \\
 & & Acc & $F_1$ & Acc & $F_1$ & Acc & $F_1$ & Acc & $F_1$ & Acc & $F_1$ & Acc & $F_1$ \\

\midrule
\multirow{3}{*}{\makecell{\textbf{PBMC} \\ \textbf{10K}}}

& Geneformer    & \ding{55} & \ding{55} & 54.0 & 42.2 & 70.3 & 46.7 & 81.0 & 63.9 & 80.9 & 71.2 & 88.0 & 78.6\\
\cmidrule{2-14}
& \textbf{LangCell $_{\bm{\mathrm{w/o CTM}}}$ } & 85.8 & 89.4 & 86.6 & \textbf{90.6} & 89.3 & 91.6 & 92.2 & 93.0 & 93.0 & 93.3 & 96.2 & 94.2 \\
& \textbf{LangCell} & \textbf{86.5} & \textbf{89.6} & \textbf{88.1} & 87.5 & \textbf{95.1} & \textbf{94.7} & \textbf{96.0} & \textbf{94.8} & \textbf{96.3} & \textbf{95.3} & \textbf{96.8} & \textbf{95.2} \\
\midrule  
\specialrule{0em}{0.5pt}{1pt}
\midrule

\multirow{3}{*}{\makecell{\textbf{PBMC} \\ \textbf{3\&68K}}} 

& Geneformer    & \ding{55} & \ding{55} & 21.1 & 24.7 & 55.2 & 49.2 & 59.3 & 69.1 & 81.5 & 74.8 & 83.3 & 74.1 \\
\cmidrule{2-14}
& \textbf{LangCell $_{\bm{\mathrm{w/o CTM}}}$ }    & 83.7 & 82.4 & \textbf{86.6} & 84.4 & 87.6 & 86.1 & 88.7 & 86.9 & 88.2 & 86.8 & 89.1 & 87.6 \\
& \textbf{LangCell} & \textbf{83.9} & \textbf{82.6} & \textbf{89.7} & \textbf{87.1} & \textbf{89.9} & \textbf{87.8} & \textbf{90.3} & \textbf{87.7} & \textbf{92.1} & \textbf{87.5} & \textbf{92.4} & \textbf{88.5}  \\
\bottomrule
\end{tabular}}
\end{table}

\section{Experiment Settings for Pre-training and the Downstream Tasks}
\label{sec:config}
\paragraph{Pre-training}
The pre-training is conducted on four NVIDIA Tesla A100 GPUs and takes approximately 50 days to complete. More experiment configurations are shown in Table \ref{Hyperparameters}.

\begin{table}[ht]
\small
    \caption{Experiment Configurations}
    \label{Hyperparameters}
    \centering
    \resizebox{0.6\textwidth}{!} {
        \begin{tabular}{lll}
        \toprule
         & \textbf{Hyperparameter} & \textbf{Value} \\
        \midrule
        Model & {\makecell[l]{Vocab size \\ Hidden size \\Number of hidden layers \\ Max sequence length \\ Number of attention heads \\ Dropout \\ Hidden act \\ LayerNorm eps}} &  {\makecell[l]{25427 \\ 512 \\ 12 \\ 2048 \\ 8 \\ 0.02 \\ ReLU \\ 1e-12}} \\
        \midrule
        Pre-training & {\makecell[l]{Similarity function \\ Optimizer \\ Scheduler \\ Max learning rate\\Warm up steps\\Weight decay\\Batch size\\Gradient accumulation}} & {\makecell[l]{Cosine similarity \\ AdamW \\ Linear \\ 1e-5\\ 1000 \\ 1e-3 \\ 3 \\ 32}} \\
        \bottomrule
        \end{tabular}
    }
\end{table}

\paragraph{Downstream Tasks} 
In downstream tasks, we uniformly follow the settings below:
\begin{itemize}
    \item Perform quality control on all datasets used, removing special categories such as ``Other'' or ``Unknown'', as well as single cells with too few expressed genes.
    \item For tasks with randomness, perform three random iterations and take the average.
    \item In few-shot tasks, all models are trained for 20 epochs.
    \item For fine-tuning tasks, all models are trained for the same number of epochs. Cell type annotation uses a training:test split of 2:1, while pathway identification uses a training:test split of 3:7.
\end{itemize}

\section{Datasets}
\label{sec:app_Datasets}
\subsection{Pre-training Data}
\label{sec:app_ptdata}

We constructed a pre-training dataset named \textbf{scLibrary} from CELLxGENE \cite{cellxgeneexplore, cellxgeneDiscover}. We obtained raw count matrices of scRNA-seq data along with their associated metadata. Our criteria for selection encompassed human cells that were analyzed using the 10X Genomics sequencing technology. We filtered out data that contained duplicates, had less than 200 expressed genes, exhibited significant gaps in metadata, or were previously utilized in other analyses. 

We employed information closely related to cell identity, such as cell type (CL), cell expression phenotype (PATO), ancestral concept system descriptions (HANCESTRO), cell anatomical information (UBERON), disease definitions (Mondo), and ontologies from the Open Biological and Biomedical Ontologies Foundry (OBO Foundry) \cite{obo}, to provide professional-level textual annotations for each single-cell.

The dataset contains textual information categorized into eight distinct cell identities, which describe single-cell sequencing data from various angles, including \texttt{Assay, Cell Type, Development Stage, Disease Information, Ethnicity of Donor, Sex of Donor, Tissue Information, Organ Information}. After pre-processing, only three labels related to the donor information—``development stage'', ``ethnicity'', and ``sex''—have missing values, while the other five labels—assay, cell type, tissue, organ, and disease—are complete without any missing values. The statistical results are depicted in Table \ref{miss_value_data_count}.

\begin{table}[ht]
\caption{Missing values in the \textbf{scLibrary}.}
\label{miss_value_data_count}
\begin{center}
\begin{tabular}{l|c}
        \toprule
        Missing Labels & Number of Cells\\
        \midrule
        Miss \texttt{Development Stage} & 1.92M\\
        Miss \texttt{Ethnicity} & 13.22M\\
        Miss \texttt{Sex} & 2.00M\\
        \midrule
        \specialrule{0em}{0.5pt}{1pt}
        \midrule
        Miss 1 label & 10.59M\\
        Miss 2 labels & 1.70M\\
        Miss 3 labels & 1.05M\\
        \bottomrule
    \end{tabular}
\end{center}
\vskip -0.1in
\end{table}

\begin{figure}[htp]
\vskip 0.2in
\begin{center}
\centerline{\includegraphics[width=0.9\columnwidth]{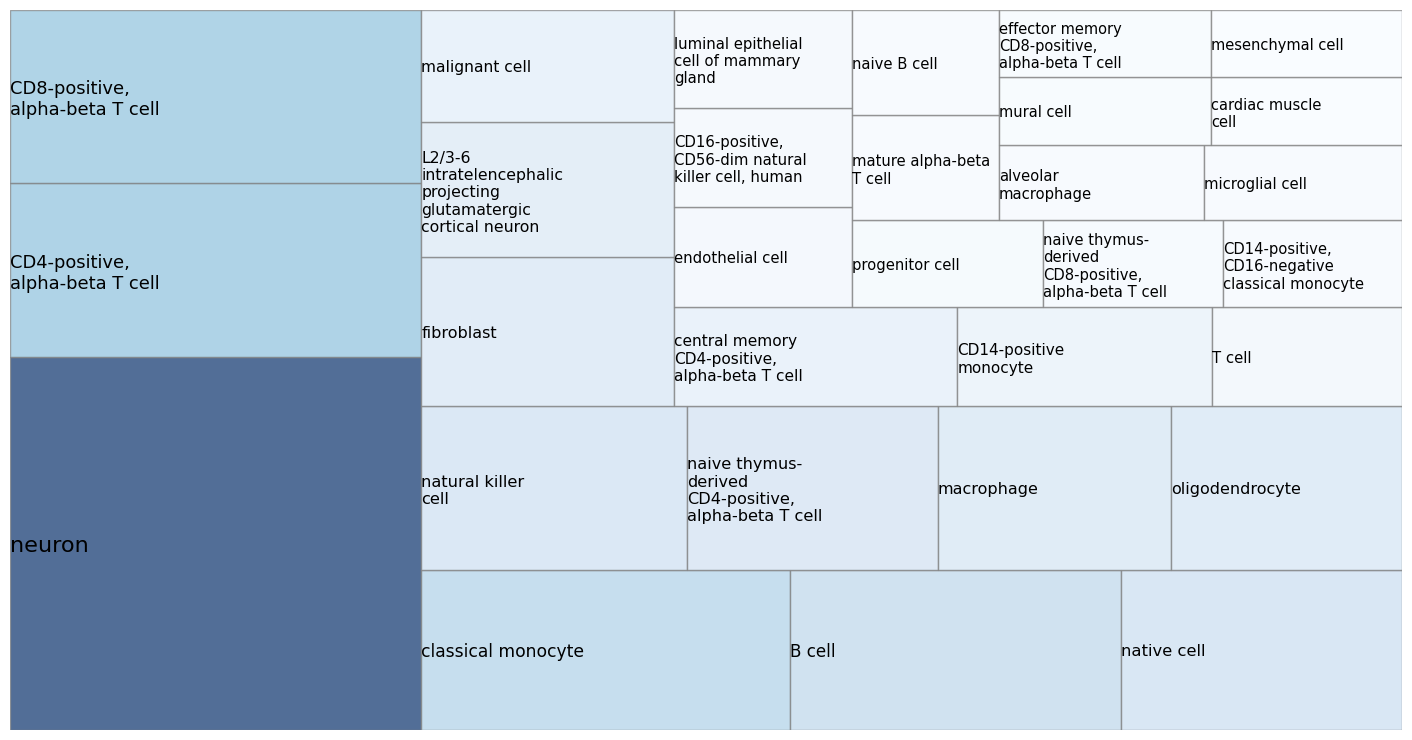}}
\caption{Overview of the distribution of cell type categories in the scLibrary dataset. To facilitate the presentation, we have selected the top 30 categories with the highest data volume for display.}
\label{obo_celltype}
\end{center}
\vskip -0.2in
\end{figure}

\begin{figure}
    \centering
    \includegraphics[width=0.9\linewidth]{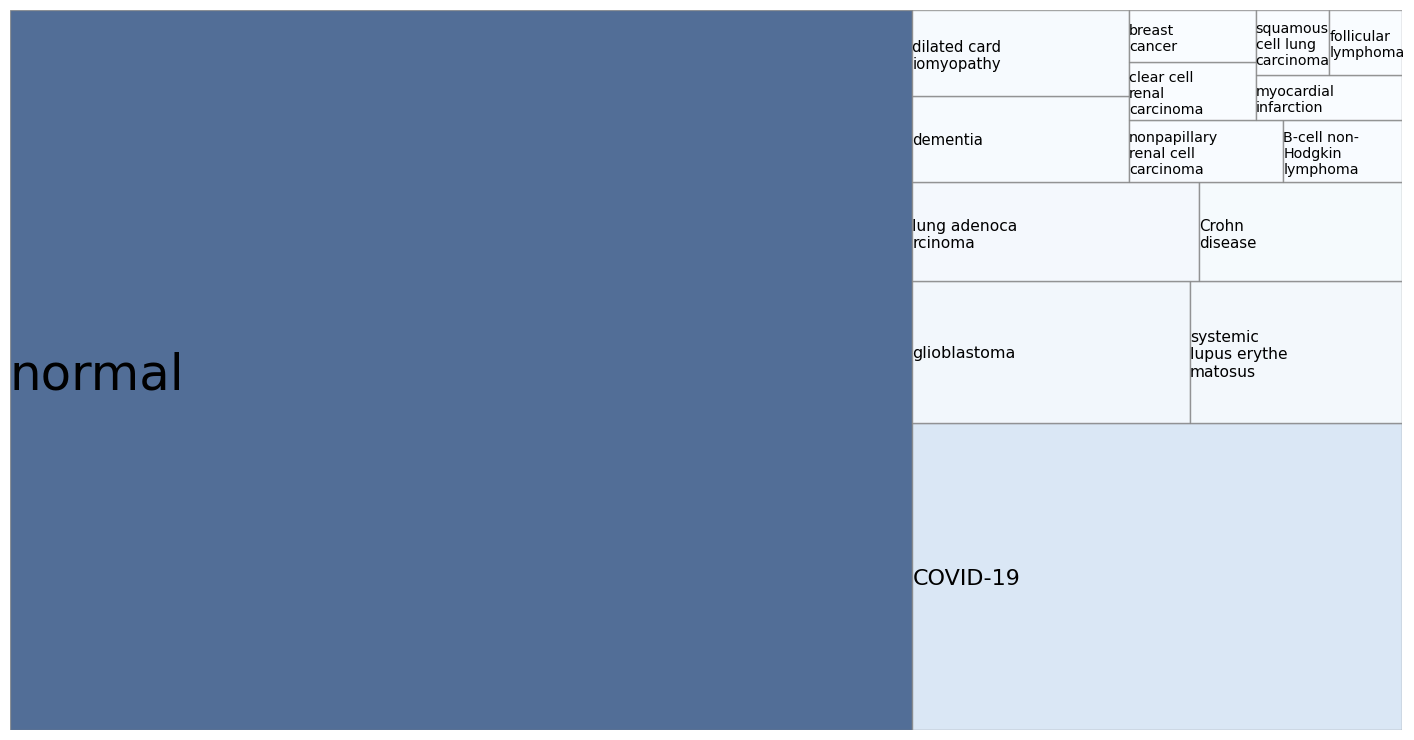} 
    \caption{Overview of the distribution of disease in the scLibrary dataset. To facilitate the presentation, we have selected the top 15 categories with the highest data volume for display.}
    \label{obo_disease}
\end{figure}

\begin{figure}
    \centering
    \includegraphics[width=0.9\linewidth]{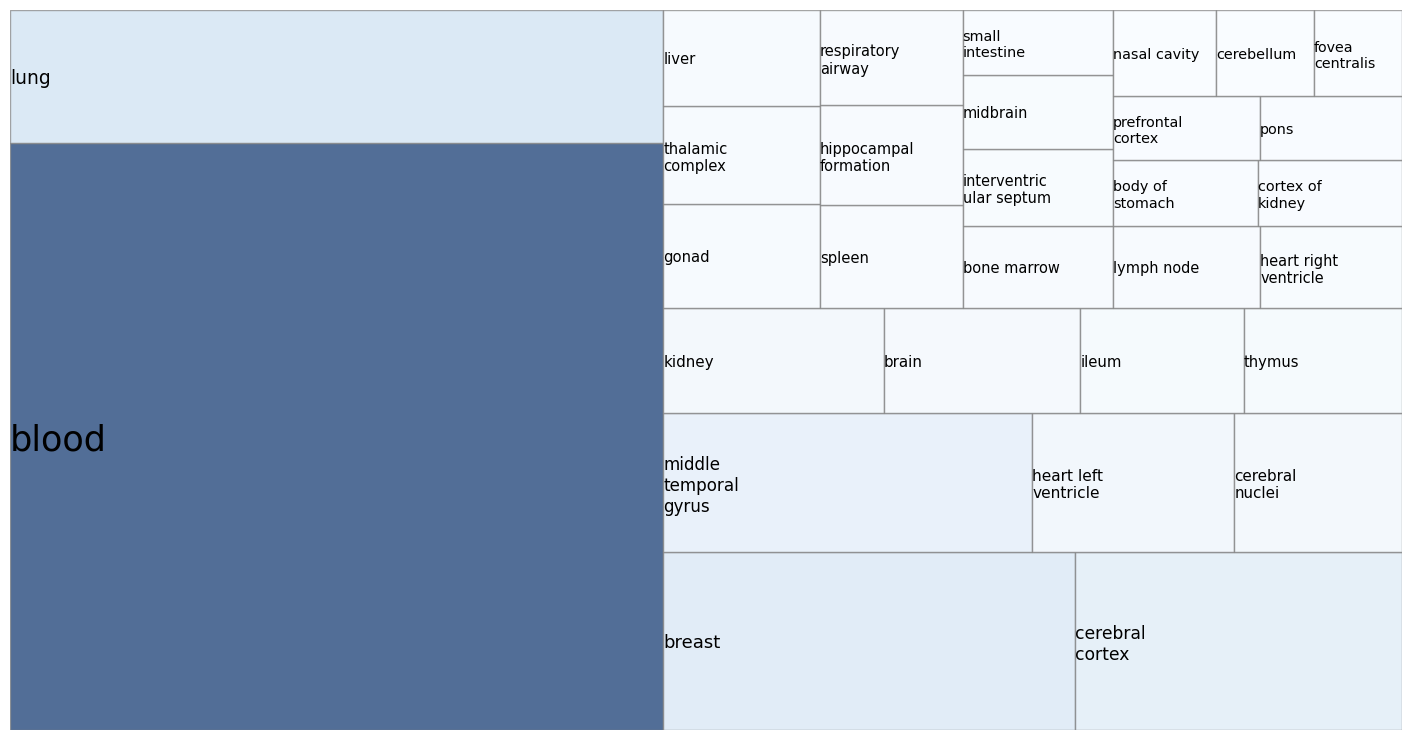} 
    \caption{Overview of the distribution of tissue in the scLibrary dataset. To facilitate the presentation, we have selected the top 30 categories with the highest data volume for display.}
    \label{obo_tissue}
\end{figure}

Each cell identity has multiple possible values (Table \ref{pre_train_data_count}). We have selected three significant cell identities to showcase the data distribution, as depicted in Fig. \ref{obo_celltype}, Fig. \ref{obo_disease}, and Fig. \ref{obo_tissue}.

\begin{table}[ht]
\caption{The cell identities used in the scLibrary.}
\label{pre_train_data_count}
\begin{center}
\begin{tabular}{l|c}
        \toprule
        Cell Identity & Number of values\\
        \midrule
        Assay & 7\\
        Disease & 56\\
        Cell Type & 562\\
        Development Stage & 160\\
        Ethnicity of Donor & 25\\
        Sex of Donor & 3\\
        Tissue Information & 192\\
        Organ Information & 48\\
        \bottomrule
    \end{tabular}
\end{center}
\vskip -0.1in
\end{table}

In the pre-training phase, we stack the cell identity information in a fixed order. Below is an example of a cell description text missing the ``ethnicity'' information:

\begin{mdframed}[backgroundcolor=blue!5!white, linecolor=blue!75!black, nobreak=true]
assay: 10x 3' v2. ;\\
cell type: malignant cell. a neoplastic cell that is capable of entering a surrounding tissue. ;\\
development stage: 74-year-old human stage. adult stage refers to an adult who is over 74 and under 75 years old. ;\\
disease: squamous cell lung carcinoma. a carcinoma arising from squamous bronchial epithelial cells. it may be keratinizing or non-keratinizing. keratinizing squamous cell carcinoma is characterized by the presence of keratinization, pearl formation, and/or intercellular bridges. non-keratinizing squamous cell carcinoma is characterized by the absence of keratinization, pearl formation, and intercellular bridges. cigarette smoking and arsenic exposure are strongly associated with squamous cell lung carcinoma. ;\\
sex: male. a biological sex quality inhering in an individual or a population whose sex organs contain only male gametes. ;\\
tissue: lung. respiration organ that develops as an out pocketing of the esophagus. ;\\
tissue general: lung. respiration organ that develops as an out pocketing of the esophagus.
\end{mdframed}

During inference, the model can work quite well even with only a single piece of identity information. For example, in the cell type annotation experiment (\ref{sec:4.2.1}), we used only the text of ``cell type''; in the NSCLC subtype classification experiment (\ref{sec:4.2.2}), we used only the text of ``disease''. Here is an example of the ``cell type'' text used in the experiment of Section \ref{sec:4.2.1}:

\begin{mdframed}[backgroundcolor=blue!5!white, linecolor=blue!75!black, nobreak=true]
cell type: dendritic cell. a cell of hematopoietic origin, typically resident in particular tissues, specialized in the uptake, processing, and transport of antigens to lymph nodes for the purpose of stimulating an immune response via T cell activation. these cells are lineage negative (cd3-negative, cd19-negative, cd34-negative, and cd56-negative).
\end{mdframed}

\subsection{Downstream Tasks Dataset}
\label{sec:app_dtdata}

We have assembled a set of benchmark datasets to evaluate the performance of LangCell across various downstream tasks. The following discussion will be structured according to the dataset of cells involved.

\paragraph{PBMC10K}

The PBMC10K dataset, as reprocessed by the study referenced in \cite{pbmc10k}, features 3,346 distinct genes that are actively expressed. It is compiled from two separate single-cell RNA sequencing (scRNA-seq) data sets, both derived from healthy human peripheral blood mononuclear cells (PBMCs). The first data set includes 7,982 individual cells, and the second comprises 4,008 cells. The PBMC10K dataset encompasses nine different cell types: B cells, CD4 T cells, CD8 T cells, CD14+ Monocytes, Dendritic cells, natural killer (NK) cells, FCGR3A+ Monocytes, Megakaryocytes, and an additional category for other cell types. We have utilized PBMC10K for zero-, few-shot, and full-data cell-type annotation tasks and single-cell integration tasks.

\paragraph{PBMC3\&68K} 

The PBMC3\&68K \cite{zheng2017massively} dataset is a comprehensive scRNA-seq dataset formed by the integration of two sub-datasets, PBMC3K, and PBMC68K, encompassing a total of 4,638 cell samples.  The dataset includes eight types of cells, which are B cells, CD4 T cells, CD8 T cells, CD14+ Monocytes, Dendritic cells, FCGR3A+ Monocytes, Megakaryocytes, and NK cells, allowing for a multifaceted exploration of cellular heterogeneity and functionality within the PBMC population. PBMC3\&68K is characterized by the analysis of 14,236 unique genes, providing a detailed view into the transcriptomic landscape of the cells. It is composed of two distinct batches, which may represent different experimental conditions or time points, offering a robust framework for comparative analysis.  This level of detail is invaluable for researchers aiming to understand the intricacies of immune cell dynamics and for the development of targeted therapeutic strategies. We have utilized PBMC3\&68K for zero-, few-shot, and full-data cell-type annotation and single-cell integration tasks.

\paragraph{Zheng68K} Zheng68K \cite{zheng2017massively} is a highly relevant and challenging dataset, consisting of 68,450 PBMCs with 11 highly related cell types. Zheng68K provides high-quality cell type annotations, making it an ideal benchmark for evaluating annotation approaches. However, the dataset poses significant challenges due to the large number of cell categories and the uneven distribution of samples between types.
We have utilized Zheng68K for zero-, few-shot, and full-data cell-type annotation tasks.

\paragraph{Human liver cross datasets}
Human liver datasets, sourced from the work \cite{scClassify}, are a combination of the macParland and aizarani datasets. The macParland dataset includes 14 cell types, while the aizarani dataset comprises 7 cell types that are part of the macParland dataset. In our experiments, we utilized the macParland dataset as the training set and the aizarani dataset as the test set to perform zero-, few-shot, and full data cell type annotation tasks, thereby assessing the model's generalization capability.

\paragraph{Perirhinal Cortex}
The original data for the Perirhinal Cortex dataset is derived from \cite{PCdataset}, which includes 606 high-quality samples from 10 distinct brain regions. The Perirhinal Cortex dataset consists of two batches with rich cellular content, containing 59,357 genes in total. The first batch includes 8,465 cells, while the second batch comprises 9,070 cells. We have utilized Perirhinal Cortex for zero-, few-shot, and full-data single-cell integration tasks.

\paragraph{Tabula Sapiens}
Tabula Sapiens \cite{Tabula_Sapiens} is an innovative human single-cell research project that has uncovered the transcriptomic features of 475 distinct cell types by analyzing live cells from multiple human tissues. The data is derived from 59 meticulously selected samples, encompassing a broad range of tissue types from the bladder to the vasculature, involving donors of varying genders, ethnicities, and ages. The project has analyzed a total of 483,152 cells, including a substantial number of immune cells, epithelial cells, endothelial cells, and stromal cells. We have utilized Tabula Sapiens for the cell-text retrieval task.

\paragraph{Non-small cell lung cancer (NSCLC) subtype dataset}

The Non-small cell lung cancer (NSCLC) subtype dataset we've developed provides a new benchmark for cell identification tasks. This dataset is sourced from CELLxGENE, where we meticulously selected cell data from the lungs of patients with malignant lung cancer, specifically those diagnosed with adenocarcinoma or squamous cell carcinoma. By annotating the dataset with clinical metadata, we've distinguished between the two NSCLC subtypes. Cluster analysis revealed a clear division of the data into two age-based clusters. Considering the more uniform data distribution in the elderly population, we have chosen this group's cluster to represent the NSCLC subtype dataset. We have utilized the NSCLC subtype dataset for the cancer subtype identification task.

\paragraph{Cell pathway identification dataset}
Pathway analysis is indispensable in this field, offering a detailed perspective on cellular diversity and molecular dynamics, which is essential for pinpointing key biological changes and therapeutic targets, thereby driving the precision of medical treatments. Therefore, we constructed the cell pathway identification dataset, a new dataset for cell identification. This dataset is obtained from CELLxGENE, and is processed using various R packages to create a pathway annotation dataset. Initially, the Seurat package was utilized for data normalization and identification of variable features. Subsequently, integrated pathway analysis was conducted on the normalized data using the irGSEA package, employing the ``AUCell'' method to score pathway activities in individual cells. The analysis specifically focused on the 50 hallmark pathways from the Broad Institute's Molecular Signatures Database (MSigDB), considering only the top 5\% of expressed pathways. Finally, the dataset was further refined to include only those pathways that appeared with a frequency greater than 0.5\%. This approach enabled a comprehensive and targeted annotation of cellular pathways, enhancing the understanding of cellular functions and states within the single-cell RNA-seq data.

It acts as a crucial lens through which we can discern cellular heterogeneity and the intricate interplay of molecular interactions within cells. By annotating active pathways in individual cells, this dataset provides an unparalleled viewpoint to identify pivotal biological transitions and pinpoint potential therapeutic targets. Such granularity is pivotal for advancing precision medicine, enabling the customization of interventions to the precise molecular characteristics observed in pathological states.

The genesis of this dataset marks the confluence of bioinformatics and systems biology, establishing a formidable foundation for forthcoming research endeavors. It enables a more profound comprehension of the molecular machinations underlying complex diseases and paves new pathways for drug discovery. By bridging the gap between high-resolution single-cell data and functional pathway analysis, this dataset emerges as a potent tool for decoding the complexities of cellular life, thereby fostering the advancement of human health and the development of innovative therapeutic strategies.

\section{Evaluation Metrics}
\label{sec:metrics}

\paragraph{Cell type annotation} 
To estimate the effectiveness of LangCell for multi-classification tasks, we employ three evaluation metrics: accuracy, macro 
$F_1$-score, and weighted $F_1$-score. Accuracy measures the closeness of the prediction to the ground truth, while macro $F_1$-score comprehensively assesses classification results without considering the importance of different categories. We also use weighted $F_1$-score to measure classification performance while accounting for the importance of different categories. These metrics are calculated based on true positive (TP), true negative (TN), false positive (FP), and false negative (FN) rates.

\begin{gather*}
    Accuracy = \frac{TP+TN}{TP+TN+FP+FN} \\
\end{gather*}
To calculate both macro $F_1$-score and weighted $F_1$-score, we need to compute Precision and Recall. These two key metrics are calculated using the following formulas:
\begin{gather*}
    Precision = \frac{TP}{TP+FP}, \quad Recall = \frac{TP}{TP+FN} \\
\end{gather*}
Thus, we can compute both macro $F_1$-score and weighted $F_1$-score using the following formulas, $N$ denotes the total number of cell types and $n_i$ denotes the number of samples in the $i$-th class:
\begin{gather*}
    macro\:F_1 = \frac{1}{N}\sum_{i=1}^{N}{F_{1}^{(i)}}\\
    weighted\:F_1 = \frac{1}{N}\sum_{i=1}^{N}{n_i*F_{1}^{(i)}} \\
    \textnormal{where} \enspace F_{1}^{(i)}=\frac{2*Precision^{(i)}*Recall^{(i)}}{Precision^{(i)}+Recall^{(i)}}
\end{gather*}

\paragraph{Single-cell Integration} 
We implemented the evaluation metrics as defined in the scIB \cite{scib} benchmark study, which serves as a benchmark for single-cell integration. Here is a detailed description of each metric.

\begin{enumerate}
    \item \textbf{Adjusted Rand Index (ARI)} 
    
    The ARI for cell types is a metric used in the field of cell biology and systems biology to evaluate the quality of cell clustering. It is a modification of the traditional Rand Index, which measures the similarity between two partitions of the same set of elements. The ARI$_{cell}$ is specifically tailored to assess the agreement between the annotated cell types (or labels) and the clusters generated by an algorithm which is a community detection algorithm applied to cell data.

    The ARI$_{cell}$ score is a normalized measure that ranges from 0 to 1. A score of 0 indicates that the clustering is no better than random chance, meaning the algorithm's partitioning is as likely as random labeling. Conversely, a score of 1 indicates a perfect match between the algorithm's clusters and the true annotations, signifying that the clustering has successfully identified the underlying structure of the cell types.

    \begin{gather*}
    RI = \frac{TP+TN}{TP+TN+FP+FN} \\
    ARI = \frac{RI- E(RI)}{max(RI)-E(RI)}
    \end{gather*}
    
    \item \textbf{Normalize Mutual Information (NMI)} 
    
    NMI is a statistical measure used to evaluate the similarity between two categorical label sets. In the single-cell integration task, we compare the ground truth cell type labels with the cell type labels derived from Louvain clustering of integrated cell embeddings. The NMI$_{cell}$ quantifies the concordance between these two sets of labels, with a score of 1 indicating perfect alignment and a score of 0 indicating no correlation. The Louvain clustering algorithm is applied across a range of resolutions from 0.1 to 2, with increments of 0.1, to find the optimal clustering configuration that maximizes the NMI$_{cell}$ score, thereby ensuring the best possible match between the predicted cell types and the actual cell types.
    \begin{gather*}
    NMI(Y, C) = \frac{2\times I(Y;C)}{H(Y)+H(C)} 
    \end{gather*}

    $Y$ represents the true categories of the data; $C$ represents the results of the clustering; $H$ represents the cross-entropy; $I(Y;C)$ represents mutual information, which is a useful measure of information in information theory. $I(Y;C) = H(Y)-H(Y|C)$. Mutual information is a useful measure of information in information theory; it represents the amount of information about one random variable contained within another, or the reduction in uncertainty of one random variable due to the knowledge of another. In other words, it quantifies the degree of correlation between two random variables.

    \item \textbf{Average Silhouette Width (ASW)} 
    
    ASW is a metric used to evaluate the quality of clustering in datasets, particularly in the context of cell type clustering and batch mixing evaluation. It quantifies the cohesion of clusters by measuring the average silhouette width of all data points within a cluster.

    The silhouette width ranges from -1 to 1, where: 
    \begin{enumerate}
        \item A value of 1 indicates that the data point is well-matched to its cluster and very dissimilar to the nearest cluster.
        \item A value close to 0 suggests that the data point lies on or near the decision boundary between two clusters, indicating poor clustering.
        \item A negative value indicates that the data point is closer to a different cluster than its own, suggesting misclassification.
    \end{enumerate} 
    
    To assess the effectiveness of cell type clustering, we calculate ASW$_{cell}$ with the known cell type labels. To evaluate batch clustering, we derive an adjusted ASW score by incorporating batch labels and subtracting 1 from it, which we refer to as ASW$_{batch}$. The scores of ASW$_{cell}$ and ASW$_{batch}$ range from 0 to 1, with higher values signifying superior performance in cell-type clustering or batch mixing. The calculation is as follows:
    \begin{gather*}
    ASW_{cell} = \frac{ASW_C + 1}{2} \\
    ASW_{batch} = 1 - |ASW_B|
    \end{gather*}
    where $C$ denotes the set of all cell identity labels.

    \item \textbf{Integration Metrics} 
    
    We report three key evaluation metrics to assess the performance of LangCell on single-cell integration tasks. Avg$_{bio}$ represents the average value of ARI$_{cell}$, NMI$_{cell}$ and ASW$_{cell}$, reflecting the conservation of biological variance. ASW$_{batch}$ indicates the effectiveness of batch effect removal. We perform a weighted average of the two to obtain S$_{final}$, providing a comprehensive evaluation of a model's performance in single-cell integration tasks.
     
    \begin{gather*}
    Avg_{bio} =  \frac{ARI_{cell} + NMI_{cell} + ASW_{cell}}{3}\\
    S_{final} =  Avg_{bio} \times 0.6 + ASW_{batch} \times 0.4
    \end{gather*}

\end{enumerate}

\paragraph{Cell-Text Retrieval}
We utilize the commonly used retrieval metric, recall@k. Specifically, for the $i$-th sample with label $y_i$, and the top $k$ results retrieved denoted as $R_{i, k}$, then:

\begin{gather*}
\text{retrieval@k}_i=1 \; if \; y_i \in R_{i, k}\; else\; 0 \\
\text{retrieval@k} = \text{average} \{\text{retrieval@k}_i\}
\end{gather*}

\paragraph{Pathway Identification}
\label{pathway}
We calculate AUROC and AUPRC in two different ways. The first method is denoted as ``avg-'', which involves calculating the AUROC and AUPRC for $N$ samples across 41 pathways separately and then taking the average. The second method is referred to as ``flatten-'', where each pathway prediction for every sample is treated as a single prediction, and the AUROC and AUPRC are computed across $41*N$ predictions.

\section{More Related Works}
\paragraph{scRNA-seq Data Representation}
The gene expression profile, essential for scientific inquiry, elucidates the intricacies of gene expression within individual cells. The presence of nearly 20,000 human protein-coding genes \cite{HGNC_Seal2022GenenamesorgTH}, coupled with the ``Dropout Zeros'' phenomenon \cite{DZ2_Svensson2019DropletSI, DZ3_Silverman2018NaughtAZ, DZ1_Linderman2022ZeropreservingIO}, significantly complicates the analysis of high-dimensional data.

Traditional methods involve dimensionality reduction, such as manual marker gene selection \cite{Pasquini2021AutomatedMF, Guo2021scSorterAC}, machine learning techniques \cite{PCA_FRS1901LIIIOL, PCA_Shen2009PrincipalCA, PCA_Hotelling1933AnalysisOA, PCA_Tsuyuzaki2019BenchmarkingPC, tsne_Maaten2014AcceleratingTU, li2023single}, or autoencoder-based approaches \cite{AE_Alessandri2020SparselyconnectedA, AE_Talwar2018AutoImputeAB, AE_Tran2019FastAP, AE_Tran2022scCANSC}. 
However, manual gene selection is often empirical \cite{problem_Huang2020EvaluationOC} and results in information loss, while machine learning methods tend to be complex and susceptible to noise. Autoencoder-based approaches depend on the similarity between test and training data. Yet, in practice, it is not always feasible to obtain labeled training data that closely match the distributions of interest.

scBERT \cite{ScBERT_yang_2022}, using the Performer architecture \cite{Performers_choromanski2022rethinking} with 6 million parameters, encodes over a million normalized, unlabeled scRNA-seq samples and surpasses performance benchmarks in cell type annotation tasks. 
Exceiver \cite{Connell_2022_neruips}, with the Perceiver IO architecture \cite{Perceiver_IO_jaegle2022perceiver}, pre-trains on 0.5 million healthy human scRNA-seq count matrix data, demonstrating effectiveness in downstream tasks. 
Geneformer \cite{geneformer} pre-trains on nearly 30 million scRNA-seq samples, applying transfer learning across various biological tasks. scGPT \cite{scGPT} is trained on over 33 million scRNA-seq records, and fine-tunes for downstream tasks including cell type annotation and multi-batch integration. scFoundation \cite{scfoundation}, with 100 million parameters, pre-trains on over 50 million human scRNA-seq data, introducing read-depth-aware pre-training to model gene co-expression patterns, validated in tasks like gene expression enhancement and drug response prediction. BioTranslator \cite{BioTranslator} bridges the gap between natural language and scRNA-seq data. However, its reliance on MLP for encoding scRNA-seq data falls short of capturing the intricacies of transcriptomic complexity.

During the review and revision process of this paper, there have also been new preprints exploring the integration of single-cell data and natural language from different perspectives. For example, Cell2Sentence \cite{levine2023cell2sentence} and GenePT \cite{chen2023genept} have proposed the idea of directly transcribing single-cell gene sequences into natural language and utilizing large language models for encoding. These works are still undergoing updates and improvements, and we look forward to their future contributions in providing more valuable insights to this field.

\paragraph{Multi-modal in Scientific Data}

Multi-modal learning enhances the model's ability to understand and express multi-modal data, with the core of this approach resting in the creation of a unified representational space that fosters inter-modal interaction and learning, capturing data interconnections and enhancing generalization through cross-modal knowledge transfer. The vision language model has experienced significant advancements. The CLIP \cite{clip} model enables image classification and description without extra supervision by learning image-text associations. BLIP \cite{Li2022BLIPBL} introduces the Multimodal Encoder-Decoder (MED) architecture, which enables the model to switch seamlessly between encoding and generation tasks, thereby enhancing the quality of the text corpus through the innovative Captioning and Filtering (CapFilt) method. Additionally, BLIP-2 \cite{Li2023BLIP2BL} leverages the Querying Transformer (Q-Former) to effectively bridge the gap between visual and textual modalities, further advancing the state of the art in vision-language pre-training.

\section{More Figures and Tables}


\begin{table}[htpb]
    \caption{The downstream tasks, categories, batch, and quantity information of each dataset used in LangCell.}
    \centering
    \resizebox{0.9\textwidth}{!} {
        \begin{tabular}{llclr}
        \toprule
        \textbf{Dataset} &
        \textbf{Downstream Task} &
        \textbf{Batch Number} &
        \textbf{Cell type} & 
        \textbf{\#Number}\\
        \midrule
        \multirow{9}{*}{\textbf{PBMC10K}} &
        \multirow{9}{*}{\makecell[l]{
        \textbf{Cell Type Annotation}\\
        (Zero-, Few-shot, and Full)\\ 
        \textbf{Single-cell Integration}}} &
        \multirow{9}{*}{\textbf{2}}
         & CD4 T cells & 4,996 \\
        & & & CD14+ Monocytes & 2,227 \\
        & & & B cells & 1,621 \\
        & & & CD8 T cells & 1,448 \\
        & & & Other & 463 \\
        & & & NK cells & 457 \\
        & & & FCGR3A+ Monocytes & 351 \\
        & & & Dendritic Cells & 339 \\
        & & & Megakaryocytes & 88 \\
        \midrule  
        \specialrule{0em}{0.5pt}{1pt}
        \midrule
        \multirow{8}{*}{\textbf{PBMC3\&68K}} &
        \multirow{8}{*}{\makecell[l]{
        \textbf{Cell Type Annotation}\\
        (Zero-, Few-shot, and Full)\\ 
        \textbf{Single-cell Integration}}} &
        \multirow{8}{*}{\textbf{2}}
         & CD4 T cells & 2,384\\
        & & & CD8 T cells & 665\\
        & & & CD14+ Monocytes & 564\\
        & & & B cells & 476\\
        & & & NK cells & 276\\
        & & & FCGR3A+ Monocytes & 195\\
        & & & Dendritic cells & 61\\
        & & & Megakaryocytes & 17\\
        \midrule  
        \specialrule{0em}{0.5pt}{1pt}
        \midrule
        \multirow{11}{*}{\textbf{Zheng68K}} &
        \multirow{11}{*}{\makecell[l] {\textbf{Cell Type Annotation}\\
        (Zero-, Few-shot, and Full)}} &
        \multirow{11}{*}{\textbf{-}}
         & CD8+ Cytotoxic T & 20,757 \\
         & & & CD8+/CD45RA+ Naive Cytotoxic & 16,645  \\
         & & & CD56+ NK & 8,775 \\
         & & & CD4+/CD25 T Reg & 6,185  \\
         & & & CD19+ B & 5,877  \\
         & & & CD4+/CD45RO+ Memory  & 3,059  \\        
         & & & CD14+ Monocyte  & 2,847 \\
         & & & Dendritic  & 2,095\\
         & & & CD4+/CD45RA+/CD25- Naive T  & 1,871  \\
         & & & CD34+   & 242\\
         & & & CD4+ T Helper2  & 97  \\
        \midrule  
        \specialrule{0em}{0.5pt}{1pt}
        \midrule
        \multirow{14}{*}{\textbf{macParland}} &
        \multirow{14}{*}{\makecell[l]{
        \textbf{Cell Type Annotation}\\
        (Zero-, Few-shot, and Full)}} &
        \multirow{14}{*}{\textbf{-}}
        & Hepatocytes & 3,501\\
        & & & ab T & 961 \\
        & & & Inflammatory Macs & 813\\
        & & & gd T & 569\\
        & & & Plasma cells & 511\\
        & & & NK cells & 488\\
        & & & Non-inflammatory Macs & 379\\
        & & & Central venous liver sinusoidal endothelial cells & 327\\
        & & & Periportal liver sinusoidal endothelial cells & 306\\
        & & & Portal endothelial cells & 211\\
        & & & Mature B cells & 129\\
        & & & cholangiocytes & 119\\
        & & & Erythroid cells & 93\\
        & & & Stellate cells  & 37 \\
        \midrule  
        \specialrule{0em}{0.5pt}{1pt}
        \midrule
        \multirow{7}{*}{\textbf{Aizarani}} & 
        \multirow{7}{*}{\makecell[l]{
        \textbf{Cell Type Annotation}\\
        (Zero-, Few-shot, and Full)}} &
        \multirow{7}{*}{\textbf{-}}
         & Hepatocytes  & 3,086 \\
        & & & T/NK & 3,066 \\
        & & & liver sinusoidal endothelial cells & 1,361 \\
        & & & cholangiocytes & 1,022  \\
        & & & macrovascular endothelial cells & 355  \\
        & & & B & 244  \\
        & & & Stellate cells and myofibroblasts &  28  \\
        \midrule
        \specialrule{0em}{0.5pt}{1pt}
        \midrule
        \multirow{10}{*}{\textbf{Perirhinal Cortex}} &
        \multirow{10}{*}{\textbf{Single Cell Integration}} &
        \multirow{10}{*}{\textbf{2}}
         & oligodendrocyte precursor cell & 6,404\\
        & & & astrocyte & 5,319\\
        & & & oligodendrocyte & 4,073\\
        & & & central nervous system macrophage & 770\\
        & & & endothelial cell & 544\\
        & & & fibroblast & 305\\
        & & & pericyte & 68\\
        & & & leukocyte & 41\\
        & & & vascular associated smooth muscle cell & 8\\
        & & & neuron & 3\\
        \bottomrule
        \end{tabular}
    }
    \label{dataset_info1}
\end{table}

\begin{table}[htpb]
    \caption{The downstream tasks, categories, batch, and quantity information of each dataset used in LangCell.}
    \centering
    \resizebox{0.9\textwidth}{!} {
        \begin{tabular}{llclr}
        \toprule
        \textbf{Dataset} &
        \textbf{Downstream Task} &
        \textbf{Batch Number} &
        \textbf{Cell type} & 
        \textbf{\#Number}\\
    
        \midrule
        \multirow{11}{*}{\makecell[l]{\textbf{Tabula Sapiens}\\(Top 10)}} &
        \multirow{11}{*}{\textbf{Cell-Text Retrieval}} &
        \multirow{11}{*}{\textbf{-}}
         & macrophage & 33,607\\
        & & & fibroblast & 31,125\\
        & & & B cell & 19,067\\
        & & & neutrophil & 16,992\\
        & & & memory B cell & 14,565\\
        & & & mesenchymal stem cell & 14,036\\
        & & & T cell  & 13,947\\
        & & & basal cell  &  12,991\\
        & & & CD4-positive, alpha-beta T cell & 12,870\\
        & & & classical monocyte &  12,746\\
       
        \midrule  
        \specialrule{0em}{0.5pt}{1pt}
        \midrule
        \multirow{2}{*}{\makecell[l]{\textbf{Non-small cell lung}\\ \textbf{cancer (NSCLC) subtype}}} &
        \multirow{2}{*}{\textbf{Cancer Subtype Identification}} &
        \multirow{2}{*}{\textbf{-}}
         & Squamous cell lung carcinoma & 1,600 \\
         & & & Lung adenocarcinoma & 1,058  \\
        
        \bottomrule
        \end{tabular}
    }
    \label{dataset_info2}
\end{table}


\end{document}